%% file: dune_decay_draft_v2.tex
\documentclass[aps,prd,superscriptaddress,nofootinbib,tighten,preprint]{revtex4}
\include{header}
\usepackage[utf8]{inputenc}
\usepackage{amsfonts}
\usepackage{amssymb}
\usepackage{amsmath}
\usepackage{graphicx}
\usepackage{hyperref}
\usepackage{placeins}
\usepackage{gensymb}
\usepackage{color}
\def\slc#1{\setbox0=\hbox{$#1$}           
    \dimen0=\wd0                                 
    \setbox1=\hbox{/} \dimen1=\wd1               
    \ifdim\dimen0>\dimen1                        
       \rlap{\hbox to \dimen0{\hfil/\hfil}}      
       #1                                        
    \else                                        
       \rlap{\hbox to \dimen1{\hfil$#1$\hfil}}   
       /                                         
    \fi}

\begin{document}

\title{A Study of Invisible Neutrino Decay at DUNE 
and its Effects on $\theta_{23}$ Measurement}

\author{Sandhya Choubey}
\email{sandhya@hri.res.in}
\affiliation{Harish-Chandra Research Institute, HBNI, Chhatnag Road, Jhunsi, Allahabad 211 019, India}
\affiliation{Department of Physics, School of
Engineering Sciences, KTH Royal Institute of Technology, AlbaNova
University Center, 106 91 Stockholm, Sweden}

\author{Srubabati Goswami}
\email{sruba@prl.res.in}
\affiliation{Physical Research Laboratory, Navrangpura, Ahmedabad 380 009, India}
\author{Dipyaman Pramanik}
\email{dipyamanpramanik@hri.res.in}
\affiliation{Harish-Chandra Research Institute, HBNI, Chhatnag Road, Jhunsi, Allahabad 211 019, India}

\begin{abstract}
\noindent 
We study the consequences of invisible decay of neutrinos in the 
context of the DUNE experiment. 
We assume that the third mass eigenstate is unstable and  decays 
to a  light sterile neutrino and a scalar or a pseudo-scalar.  
We consider DUNE running in 5 years neutrino and 5 years antineutrino 
mode and a detector volume of 40 kt. 
We obtain the expected sensitivity 
on the rest-frame life-time $\tau_3$ normalized to the mass
$m_3$ as   
$\tau_3/m_3 > 4.50\times 10^{-11}$ s/eV at 90\% C.L. for a normal 
hierarchical mass spectrum. 
We also find that  
DUNE can discover neutrino decay for  
$\tau_3/m_3 > 4.27\times 10^{-11}$ s/eV
at 90\% C.L. 
In addition, for an unstable $\nu_3$
with an illustrative  value of $\tau_3/m_3$ = $1.2 \times 10^{-11}$ s/eV, the no decay
case could get disfavoured at the  $3\sigma$ C.L. At 90\% C.L. the expected precision range
for this true value is obtained as  
$1.71 \times 10^{-11} > \tau_3/m_3 > 9.29\times 10^{-12}$ in units of s/eV.
We also study the correlation between a non-zero $\tau_3/m_3$  
and standard oscillation parameters and find an interesting correlation 
in the appearance and disappearance channels with the mixing angle $\theta_{23}$. 
This alters the octant sensitivity of DUNE, favorably (unfavorably) for true 
$\theta_{23}$ in the lower (higher) octant. The effect of a decaying neutrino
does not alter the hierarchy or CP violation discovery sensitivity of DUNE 
in a discernible way. 
\end{abstract}

\maketitle
\section{Introduction}

Neutrino oscillation experiments have established that neutrinos 
are massive and  there is mixing among all the three  flavors. 
The neutrino oscillation parameters -- namely the two mass squared 
differences $\Delta m^2_{21} = m_2^2 - m_1^2$ and $|\Delta m^2_{31}|= |m_3^2 - m_1^2|$, $m_1,m_2,m_3$ being the mass states   
and the three mixing 
angles $\theta_{12}, \theta_{13}$ and $\theta_{23}$ have been 
determined with considerable precision by global oscillation analysis 
of world neutrino data  \cite{Capozzi:2016rtj,Esteban:2016qun}. 
The unknown oscillation parameters are:
(i) the neutrino mass hierarchy i.e whether $m_3 > m_2 > m_1$ 
leading to normal hierarchy (NH) or if $m_3 < m_2 \approx m_1$ 
resulting in inverted hierarchy (IH); 
(ii) the octant of $\theta_{23}$ -- if $\theta_{23} < 45^\circ$ 
it is said to be in the lower octant (LO)  and if $\theta_{23} > 45^\circ$ 
it is in the higher octant (HO);
(iii) the CP phase $\dcp$. 
There are some indications of these quantities from the 
recent results of the ongoing experiments T2K \cite{Abe:2017uxa}  and
NO$\nu$A \cite{Adamson:2016tbq, Adamson:2017gxd}. However no 
clear  conclusions have been reached yet.    
While more statistics from these experiments will be able to 
give some more definite ideas regarding the three unknowns, the 
precision determination is expected to come from the  Fermilab based 
broadband experiment DUNE 
which has a baseline of 1300 km.  With a higher baseline than T2K and 
NO$\nu$A, DUNE is more sensitive to the  earth matter effect 
which gives it an edge to determine the mass hierarchy and octant 
and hence $\dcp$ by resolving the wrong-hierarchy/wrong-CP 
or wrong-octant/wrong-CP solutions.
Thus, it is expected that  
DUNE will be able to determine the above three 
outstanding parameters with considerable precision with its projected run. 
For a recent analysis, in view of the latest NO$\nu$A results see 
\cite{ Goswami:2017hcw}. 

This capability of DUNE, to make precision measurements also makes it 
an ideal experiment to test sub-dominant new physics effects. 
One of the widely studied subdominant effect, in the context
of neutrino experiments, is the ``invisible decay" of neutrinos {\it i.e.} 
decay to sterile states.  
This can arise  if  neutrinos are:  
(i) Dirac 
particles and  there is a coupling between the
neutrinos and a light scalar boson \cite{Acker:1991ej,Acker:1993sz}. 
This allows the decay 
$\nu_j \rightarrow \bar\nu_{iR} + \chi$ 
where $\bar\nu_{iR}$ is a right-handed singlet and 
$\chi$ is an iso-singlet scalar carrying a lepton number. 
(ii) Majorana particles with a pseudo-scalar coupling with a Majoron 
\cite{Gelmini:1980re,Chikashige:1980ui} 
and a sterile neutrino giving rise to the decay 
$\nu_j \rightarrow {\nu_s} + J$ . 
The Majoron should be dominantly singlet to comply with the 
constraints from LEP data on the Z decay to  invisible particles. 
Such models have been discussed in 
\cite{Pakvasa:1999ta}.  
In both the cases  
all the final state particles 
are sterile and invisible.  
The other possibility of ``visible decay'' occurs if neutrinos 
decay to active neutrinos 
\cite{Kim:1990km,Acker:1992eh,Lindner:2001fx}. 
The decay modes are
$\nu_j \rightarrow \bar{\nu_i} + J$ or $\nu_j \rightarrow \nu_i +J$.  
For Majorana neutrinos both    
$\bar{\nu}_i$ ($\nu_i$) can interact as a flavor state
$\bar{\nu}_\alpha$(${\nu_\alpha}$)
according to $U_{\alpha i}^2$ and thus the
decay product can be observed in the detector.
For recent studies on implications of visible decay 
for long baseline experiments see \cite{Coloma:2017zpg,Gago:2017zzy}.  

Neutrino decay manifests itself through a term of the form 
$\exp(-\frac{m_i L}{\tau_i E})$ which gives the fraction of neutrinos of 
energy $E$
that decays after travelling through a distance $L$. 
Here, $\tau_i$ denotes the rest frame life time of the 
state with mass $m_i$. 

Neutrino decay as a solution to explain the depletion of the 
solar neutrinos has been considered very early \cite{ Bahcall:1972my}. 
Subsequently, many authors have studied the neutrino 
oscillation plus decay solution to the  solar neutrino 
problem  and put bounds on the lifetime 
of the unstable state
\cite{Berezhiani:1991vk,Berezhiani:1992xg,Acker:1993sz,Choubey:2000an,Bandyopadhyay:2001ct,Joshipura:2002fb,Bandyopadhyay:2002qg,Picoreti:2015ika}.   
These studies 
considered $\nu_2$ to be the unstable 
state and obtained bounds on $\tau_2$. 
This is plausible, since, $U_{e3}$ being relatively small, the admixture of 
$\nu_3$ state in the solar $\nu_e$ can be taken to be small. 
Note that most of these studies have been performed prior to 
the discovery of non-zero $\theta_{13}$. 
Typical bounds obtained from solar neutrino data is 
$\tau_2/m_2 > 8.5 \times 10^{-7}$ s/eV \cite{Bandyopadhyay:2002qg}. 
A recent study performed in 
ref. \cite{Berryman:2014qha} has obtained bound on  both $\tau_2$ and 
$\tau_1$ including  low energy solar neutrino data.   
Bound on $\tau_2$ or $\tau_1$ can also come from supernova neutrinos. 
The bound obtained in ref.~\cite{Frieman:1987as} is $\tau/m > 10^{5}$ s/eV 
from SN1987A data.  

Bounds on the $\nu_3$ lifetime have been obtained in the context of 
atmospheric and long-baseline (LBL) neutrinos. 
A pure neutrino decay solution (without including neutrino 
mixing into account)  to the atmospheric neutrino problem 
was discussed in \cite{LoSecco:1998cd} 
and it was shown that this solution gives a poor fit to the data. 
Neutrino decay with non-zero neutrino mixing angle was first 
considered in \cite{Barger:1998xk,Lipari:1999vh} where it was assumed that 
the state $\nu_\mu$  has an unstable component. 
It was  shown that the 
neutrino decay with mixing can reproduce the $L/E$ distribution of the 
SuperKamiokande (SK) data. 
Later this scenario was reanalyzed in \cite{Fogli:1999qt} using the 
zenith angle dependence of the SK data rather than the $L/E$ dsitribution 
and it 
was shown that this gives a poor fit to the SK data. 
In \cite{Barger:1998xk} and \cite{Fogli:1999qt}  the $\Delta m^2$ dependent 
terms were averaged out since it was assumed that $\Delta m^2 > 0.$1 eV$^2$ 
to comply with the constraints coming from K decays \cite{Barger:1998xk}. 
However, these constraints can be evaded if the unstable state decays 
to some state with which it does not mix.
Two scenarios have been considered in the literature in this context. 
The first analysis was done in 
ref. \cite{Choubey:1999ir}, where  
$\Delta m^2$  was kept  
unconstrained and it  explicitly appeared in the probabilities. 
It was shown that 
if one fits SK data with 
this scenario then the  best-fit comes with a non-zero value of the decay parameter  
and $\Delta m^2 \sim 0.003$ eV$^2$, 
implying that oscillation combined with small decay can give a better fit 
to the data.
Later in ref. \cite{Barger:1999bg}  the possibility that $\Delta m^2 << 10^{-4}$ eV$^2$ 
was considered. In this case also the probability does not contain $\Delta m^2$ 
explicitly and the scenario correspond to decay plus mixing.
In \cite{Barger:1999bg} it was shown that this scenario can give a 
good fit to the SK data but subsequent analysis by the SK collaboration 
revealed that the decay plus mixing scenario of \cite{Barger:1999bg} 
gives a poorer fit to data than only oscillation  \cite{Ashie:2004mr}. 
A global analysis of atmospheric and long-baseline neutrino data in the framework of  
the oscillation plus decay scenario considered in \cite{Choubey:1999ir} 
was done in \cite{GonzalezGarcia:2008ru}. 
Although only oscillation gave the best- fit to the SK data, a reasonable fit 
was also obtained with oscillation plus decay.  However, with the addition of 
the data from the LBL experiment MINOS, the quality of the fit became poorer. 
They put a bound on $\tau_3/m_3 \geq 2.9 \times 10^{-10}$ s/eV at the 90\% C.L. 
Subsequently, analysis of oscillation plus decay scenario with 
unconstrained $\Delta m^2$ 
have been performed in \cite{Gomes:2014yua} in the context of MINOS and 
T2K data  
and the bound $\tau_3/m_3 > 2.8 \times 10^{-12}$ s/eV at 90\% C.L. was obtained.
\footnote{Currently NO$\nu$A and T2K are taking data. It will be interesting to see the  bounds from these experiments in the context of neutrino decay \cite{Pramanik}.}
Most of the above analyses have been performed in the two generation limit
and  matter effect in presence of decay, 
in the context of atmospheric and LBL neutrinos 
have not been studied. 
Expected constraint on  $\tau_3/m_3 > 7.5 \times 10^{-11}$ s/eV (95\% C.L.) 
in the context of the medium baseline reactor neutrino experiment like JUNO  
has been obtained in ref.  \cite{Abrahao:2015rba} using 
a full three generation picture in vacuum. 

Decay of ultrahigh energy astrophysical neutrinos 
have been considered by many authors
\cite{Beacom:2002vi,Maltoni:2008jr,Pakvasa:2012db,Pagliaroli:2015rca}.
A recent study performed in ref. \cite{Bustamante:2016ciw} 
states that the IceCube experiment 
can reach a sensitivity of 
$\tau/m \geq 10$ s/eV for both 
hierarchies for 100 TeV neutrinos coming from a source at a distance of 1 Gpc.

In this paper we study the constraints on $\tau_3/m_3$ from  simulated 
data of the DUNE experiment assuming 
invisible decay of the neutrinos. We perform a full three flavor study 
and include the matter effect during the  propagation of the neutrinos 
through the Earth.   
We obtain constraints on the life time of the unstable state 
from the simulated DUNE data. For obtaining this we assume  
the state $\nu_3$ to be unstable and obtain the 
constraints on $\tau_3/m_3$.  We also study the discovery potential 
and precision of measuring $\tau_3/m_3$ at  DUNE assuming decay is 
present in nature. 
In addition, we explore how the decay lifetime $\tau_3$ is correlated with 
the mixing angle $\theta_{23}$ leading to approximate degeneracy between 
the two parameters. We will study how this affects the sensitivity of DUNE to 
the mixing angle $\theta_{23}$ and its octant. 
The sensitivity of DUNE, to determine the 
standard oscillation parameters like the mass hierarchy and 
$\dcp$ could also in principle  get affected if the heaviest state is 
unstable, and we study this here. 

The plan of the paper is as follows. In the next section we 
discuss the propagation of neutrinos through earth matter 
when one of the state undergoes decay to invisible states. 
Section III contains the experimental and simulation details. 
The results are presented in the section IV and finally 
we conclude in section V.

\section{Neutrino  Propagation in Presence of Decay }

Assuming the $\nu_3$  state to decay into a sterile neutrino 
and a singlet scalar ($\nu_3 \rightarrow \bar\nu_4 + J$), 
the flavor and mass eigenstates get related as, 
\begin{equation}\label{mixing}
\begin{pmatrix}
\nu_\alpha\\
\nu_s
\end{pmatrix}
= 
\begin{pmatrix}
U & 0 \\
0 & 1  
\end{pmatrix}
\begin{pmatrix}
\nu_i\\
\nu_4
\end{pmatrix}. 
\end{equation}
Here, $\nu_\alpha$ with $\alpha =e,\mu,\tau$ denote the flavor states 
and $\nu_i$ , $i=1,2,3$ denote the three mass states corresponding to the 
active neutrinos. $U$ is the standard $3\times3$ PMNS mixing matrix.  
We assume that  $m_3 > m_2 > m_1 > m_4$ for NH while 
$m_2 \approx m_1 > m_3 > m_4$ for IH. 
This implies that for both NH and IH the third mass state can decay 
to a lighter sterile state.  
Note that $\Delta m^2_{34}$ is unconstrained from meson decay bound 
since $m_4$ is the mass of a  sterile state with which the active neutrinos do not mix. 
The effect of decay can be incorporated in the propagation equation by 
introducing a term  
$m_3/\tau_3$ in the evolution equation: \footnote{This implicitly assumes that the neutrino mass matrix and the decay matrix can be simultaneously 
diagonalised and the mass eigenstates are the same as the decay eigenstates
\cite{Berryman:2014yoa,Frieman:1987as}.} 
\begin{equation}\label{evol}
i\frac{d}{dx}\begin{pmatrix}
\nu_e\\
\nu_\mu\\
\nu_\tau
\end{pmatrix}
 = \left[U\left[\frac{1}{2E}
\begin{pmatrix}
0&0&0\\
0&\Delta m^2_{21}&0\\
0&0&\Delta m^2_{31}
\end{pmatrix}
-i\frac{m_3}{2 E \tau_3}
\begin{pmatrix}
0&0&0\\
0&0&0\\
0&0&1
\end{pmatrix}\right]U^\dagger
+ 
\begin{pmatrix}
      A&0&0\\0&0&0\\0&0&0
     \end{pmatrix}
\right]
\begin{pmatrix}
\nu_e\\
\nu_\mu\\
\nu_\tau
\end{pmatrix}
\,,
\end{equation}
where 
the term $A = 2 \sqrt{2} G_F n_e E$ represents the matter potential
due to neutrino electron scattering in matter, $G_F$ denotes the Fermi 
coupling, $E$ is the energy, and $n_e$ the electron density. 
For the antineutrinos the matter potential comes with a negative sign. 
The fourth sterile state 
being decoupled from the other states except through the interaction 
which induces the decay, does not affect the propagation. 
We solve the 
Eq.~\eqref{evol}, in matter numerically using Runge-Kutta technique 
assuming the PREM density profile 
\cite{prem} for the Earth matter. 
In section IV we will show the plots of the appearance channel 
probability $P_{\mu e}$ and disappearance channel probability 
$P_{\mu \mu}$ as a function of neutrino energy as well as a function of $\theta_{23}$ for a fixed neutrino energy. 

\section{Experiment and Simulation Details}

DUNE (Deep Underground Neutrino Experiment) is a long baseline neutrino oscillation experiment  proposed to be built in USA. DUNE will consist of a neutrino beam from Fermilab to Sanford Underground Research Facility \cite{Acciarri:2016crz,Acciarri:2015uup,Strait:2016mof,Acciarri:2016ooe}. The source of the beam will consist of a 80-120 GeV proton beam which will give the neutrino beam eventually. There will be a 34-40 kt liquid Argon time projection chamber (LArTPC) at the end of the beam, at 1300 km from the source. We use the GLoBES package \cite{Huber:2004ka,Huber:2007ji} to simulate the DUNE experiment. In our simulation, we have used neutrino and antineutrino flux from a 120 GeV, 1.2 MW proton beam at Fermilab and a 40 kt far detector. All our simulations are for a runtime of 5+5 years (5 years in $\nu$ and 5 years in $\bar\nu$) and we take  both appearance and disappearance channels into account, unless otherwise stated. For the charged current (CC) events we have used $20\%$ energy resolution for the $\mu$ and $15\%$ energy resolution for the $e$.

 The primary background to the electron appearance and muon disappearance signal  events come from the neutral current (NC) background and the intrinsic $\nu_e$ contamination in the flux.  In addition, there is a wrong-sign component in the flux. The problem of wrong-sign events is severe for the antineutrino run, since the neutrino component in the antineutrino flux is larger than the antineutrino component in the neutrino flux. We have taken all these backgrounds into account in our analyses. In order to reduce the backgrounds, several cuts are taken. The effects of these cuts have been imposed in our simulation through efficiency factor of the signal. We have included 2\% (10\%) signal (background) normalization errors and 5\% energy calibration error as systematic uncertainties. The full detector specification can be found in Ref. \cite{Acciarri:2015uup} and we have closely followed it. A final comment on our treatment of the NC backgrounds is in order. Since neutrino decay is a non-unitary phenomenon, we expect that the NC events at DUNE would be impacted depending on the decay lifetime. Hence, one would expect that the NC background too would depend on the decay lifetime. However, we have checked that for standard three-generation oscillations, the NC background is 71 events out of a total of 347 background events. In presence of decay ($\tau_3/m_3 = 1.2\times 10^{-11}$ s/eV), the NC background changes to 65, while the total background is 338. This implies that the NC background changes by about 8.5\% only. We have also checked that including decay in the NC background changes the $\chi^2$ only in the first place in decimal and is not even visible in the plots. Hence, for simplicity we keep the NC background fixed at the standard value.

Throughout this paper we use the following true values for the standard oscillation parameters: $\theta_{12} = 34.8\degree$, $\theta_{13} = 8.5\degree$, 
$\Delta m^2_{21} = 7.5\times10^{-5}$ eV, $\Delta m^2_{31} = 2.457\times10^{-3}$ eV and $\delta_{CP} = -90\degree$. The true value of $\theta_{23}$ will be 
specified for every plot that we will present. 
We marginalize our $\chi^2$ over $\theta_{13}$, $\theta_{23}$,  $\dcp$ and $\ma$, which are allowed to vary within their allowed $3\sigma$ ranges. We add a Gaussian prior on $\theta_{13}$ with the current $1\sigma$ range on this parameter. 
We keep the hierarchy normal in the simulated data for all the results presented in this paper. In this paper we have used ``true" to specify the parameters used for generating the simulated data and ``test" to specify the parameters in the fit.

\section{Results}

\begin{figure}[]
\begin{center}
\includegraphics[width=0.45\textwidth]{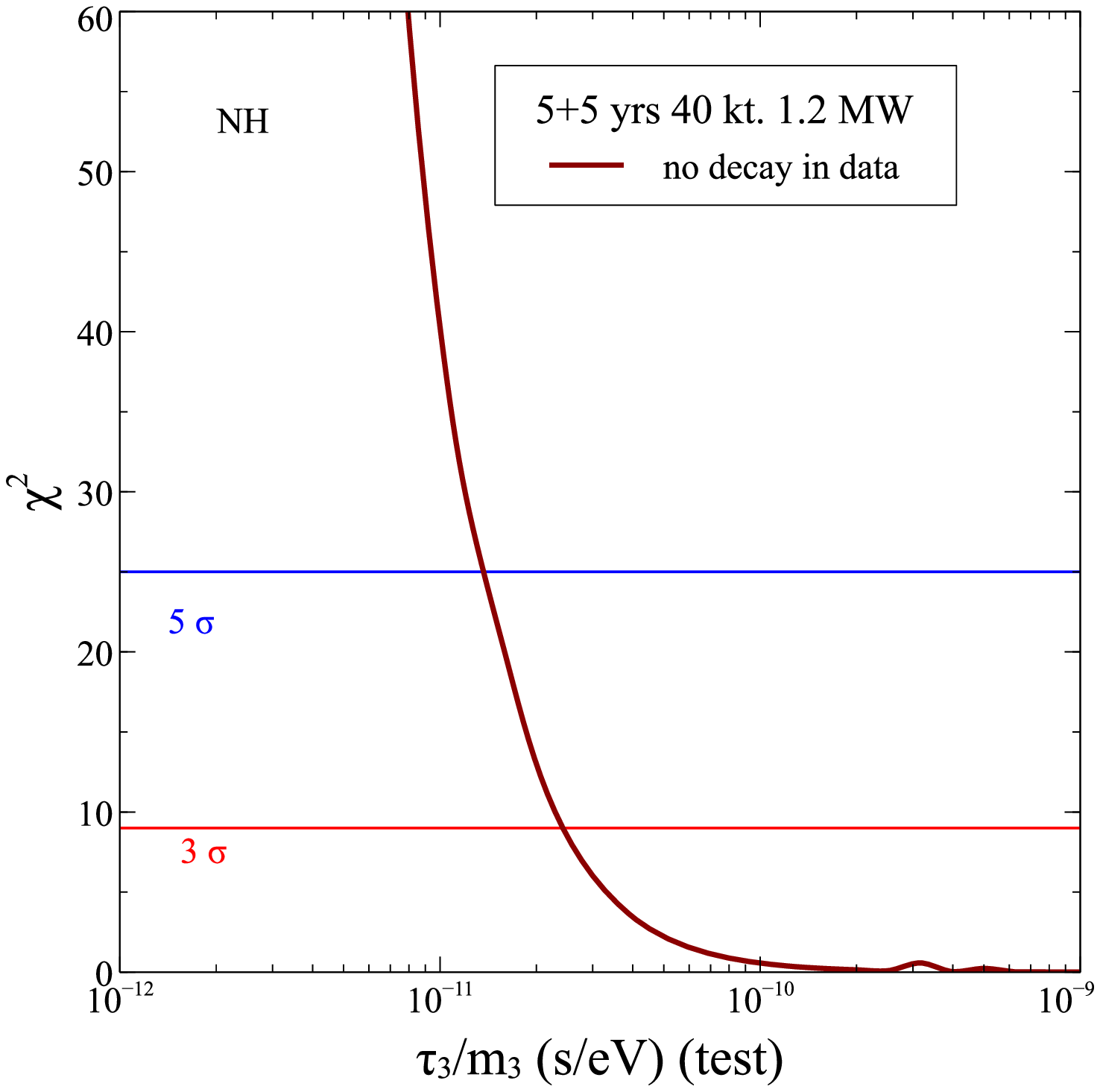} 
\includegraphics[width=0.45\textwidth]{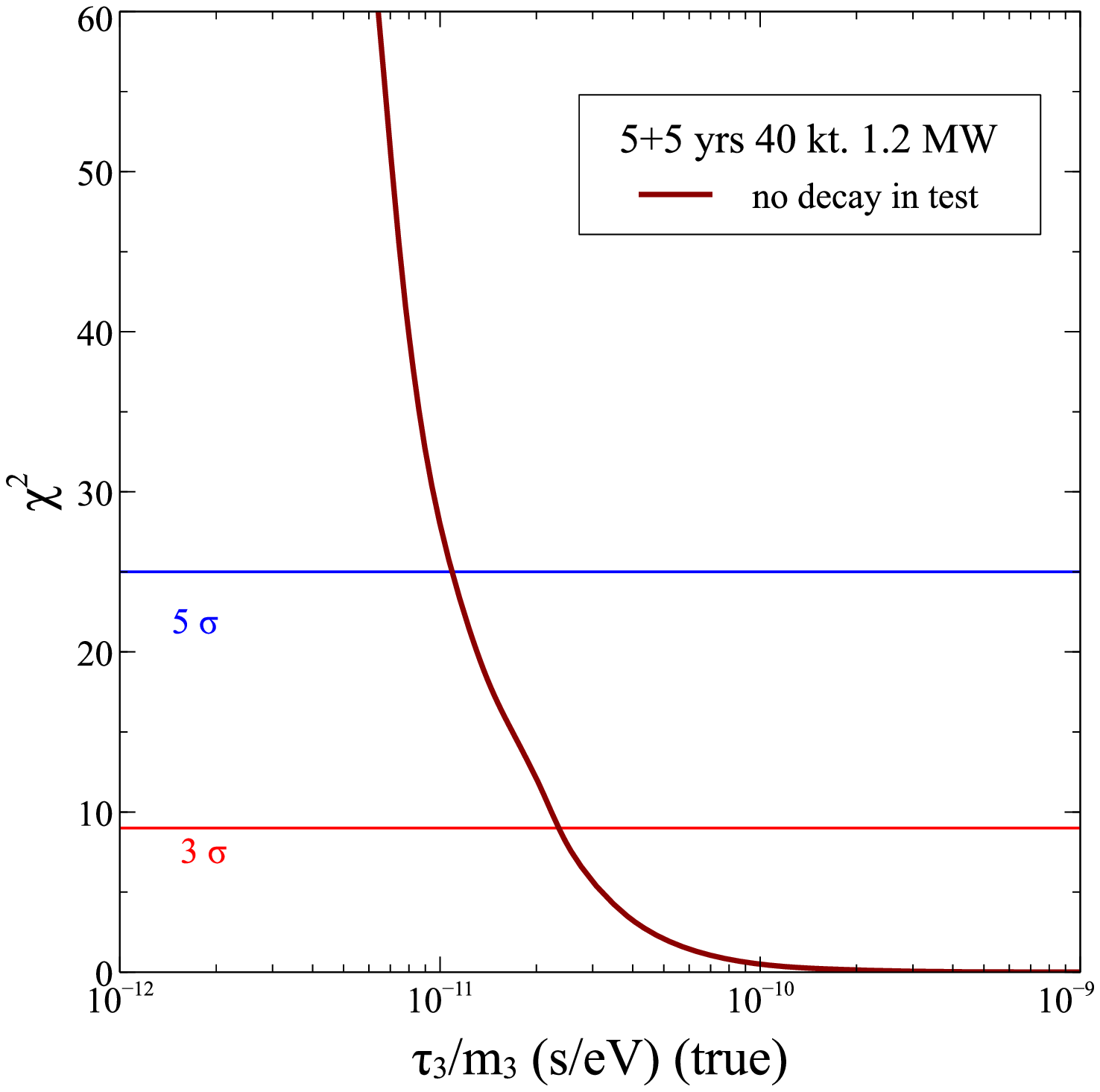}
\caption{\label{fig:sens}The left panel of this figure shows the expected sensitivity of DUNE to 
constraining the decay parameter $\tau_{3}/m_{3}$. The right panel  of this figure presents the potential of DUNE to discover a decaying $\nu_3$. See text for details. }
\end{center}
\end{figure}

We will first present results on how well DUNE will be able to constrain or discover the lifetime of the unstable $\nu_3$ state. We will then turn our focus on how the measurement of standard oscillation physics at DUNE might get affected if $\nu_3$ were to be unstable.

\subsection{Constraining the $\nu_3$ lifetime}

The left panel of Fig.~\ref{fig:sens} shows the potential of DUNE to constrain the lifetime of $\nu_3$ normalized to its mass $m_3$. In order to obtain this curve we generate the data for a no-decay case and fit this data for an unstable $\nu_3$. The data was generated at the values of oscillation parameters given in section III and $\theta_{23}=42^\circ$. We marginalize the $\chi^2$ over all standard oscillation parameters as mentioned above. We see that at the $3\sigma$ level DUNE could constrain $\tau_3/m_3 > 2.38\times 10^{-11}$ s/eV, whereas at 90\% C.L. the corresponding expected limit would be $\tau_3/m_3 > 4.50\times 10^{-11}$ s/eV. This can be compared with the current limit on $\tau_3/m_3 > 2.8\times 10^{-12}$ s/eV that we have from combined  MINOS and T2K analysis \cite{Gomes:2014yua}. Therefore, DUNE is expected to improve the bounds on $\nu_3$ lifetime by at least one order of magnitude. \footnote{Note that by the time DUNE will be operative, the current experiments would have improved their statistics and hence bounds on standard and non-standard parameters. Even then after the full run of these experiments, DUNE will have more sensitivity compared to the full run of NO$\nu$A and T2K \cite{Pramanik}.}
The right panel  of Fig.~\ref{fig:sens} shows the discovery potential of a decaying neutrino at DUNE. To obtain this curve we generated the data taking a decaying $\nu_3$ into account and fitted it with a theory for stable neutrinos. True value of $\theta_{23}=42^\circ$ for this plot and the other oscillation parameters are taken as before. The $\chi^2$ so obtained is then marginalized over the standard oscillation parameters in the fit. The resultant marginalized $\chi^2$ is shown in the right panel  of Fig.~\ref{fig:sens} as a function of the $\tau_3/m_3$(true). DUNE is expected to discover a decaying neutrino at the $3\sigma$ C.L. for $\tau_3/m_3 > 2.38\times 10^{-11}$ s/eV and the 90\% C.L. for $\tau_3/m_3 > 4.27\times 10^{-11}$ s/eV. 

\begin{figure}[]
\begin{center}
\includegraphics[width=0.45\textwidth]{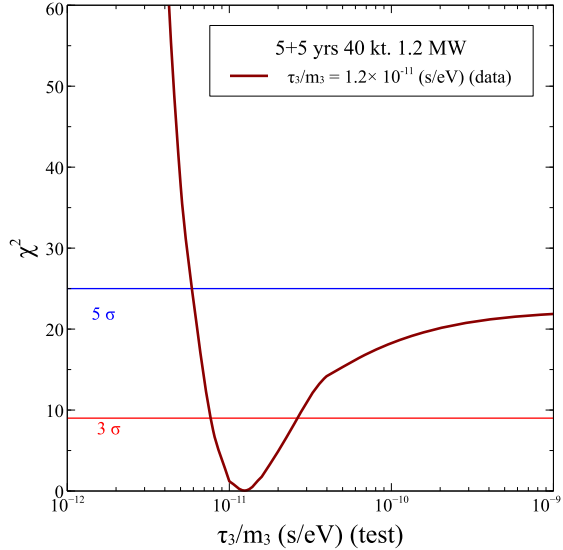} 
\caption{\label{fig:precision}The $\chi^2$ as a function of $\tau_{3}/m_{3}$(test), showing the precision with which $\tau_3/m_3$ could be measured at DUNE when its true value is $1.2\times 10^{-11}$ s/eV. }
\end{center}
\end{figure} 

Assuming that the $\nu_3$ is indeed unstable with a decay width corresponding to $\tau_3/m_3=1.2\times 10^{-11}$ s/eV, we show in Fig.~\ref{fig:precision} how well DUNE will be able to constrain the lifetime of the decaying $\nu_3$.  It is seen from the figure that in this case not only can DUNE exclude the no decay case above 3$\sigma$, but can also measure the value of the decay parameter with good precision. The corresponding $3\sigma$ and 90\% C.L. ranges are $2.63 \times 10^{-11} > \tau_3/m_3 > 7.62\times 10^{-12}$ and  $1.71 \times 10^{-11} > \tau_3/m_3 > 9.29\times 10^{-12}$ in units of s/eV, respectively.

\subsection{Constraining $\theta_{23}$ and its Octant}

\begin{figure}[]
\begin{center}
\includegraphics[width=0.45\textwidth]{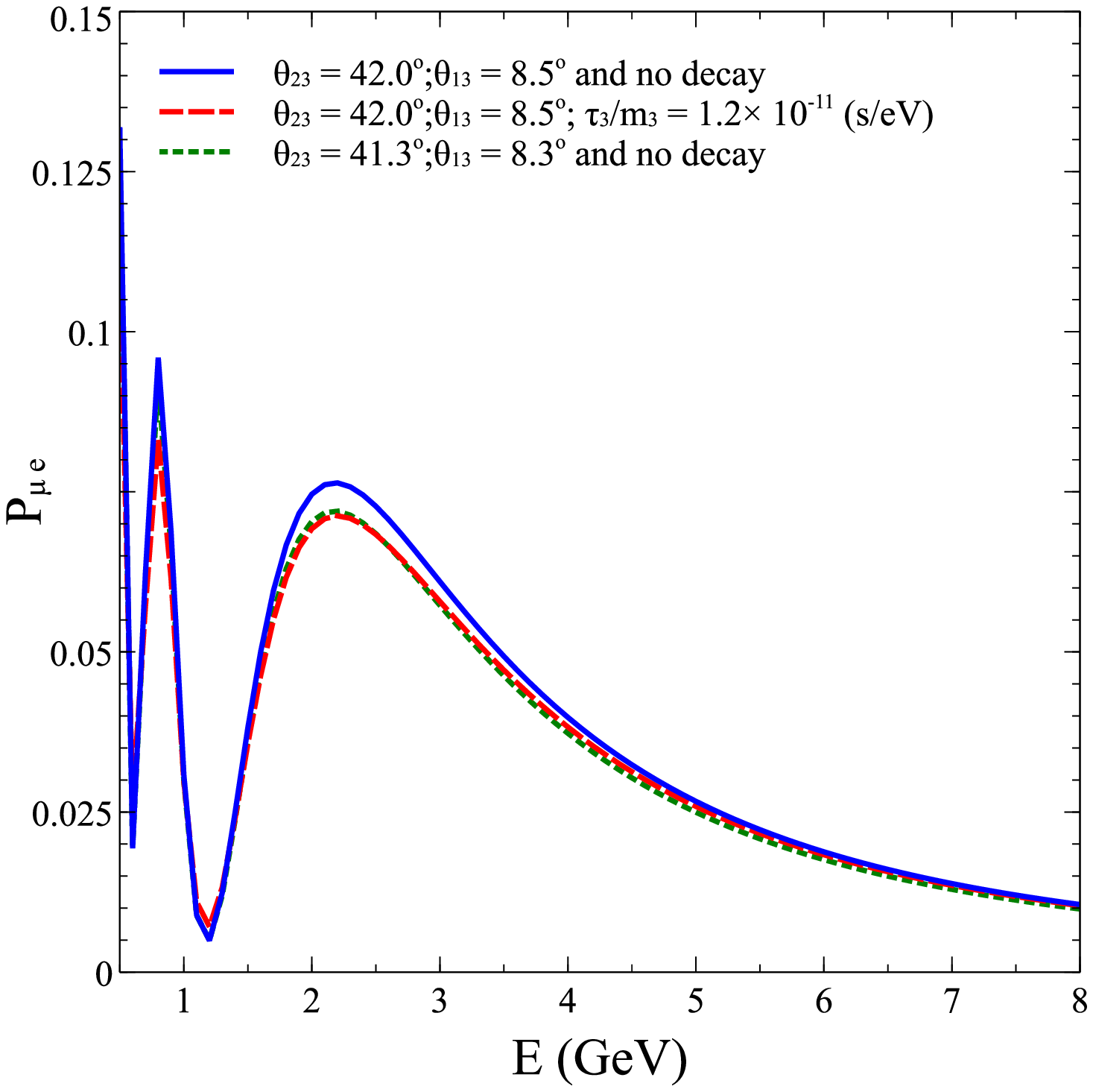}
\includegraphics[width=0.45\textwidth]{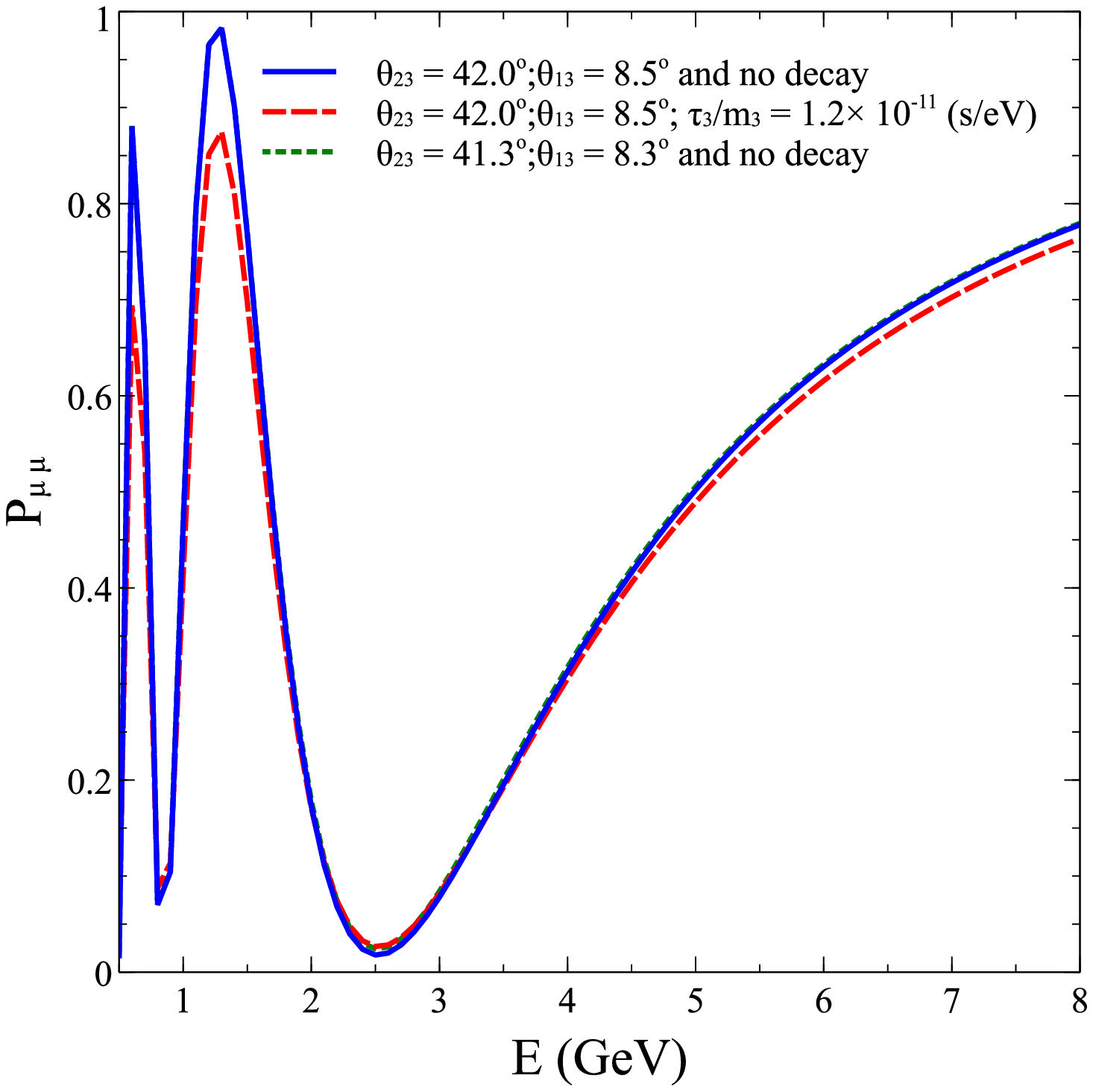}
\includegraphics[width=0.45\textwidth]{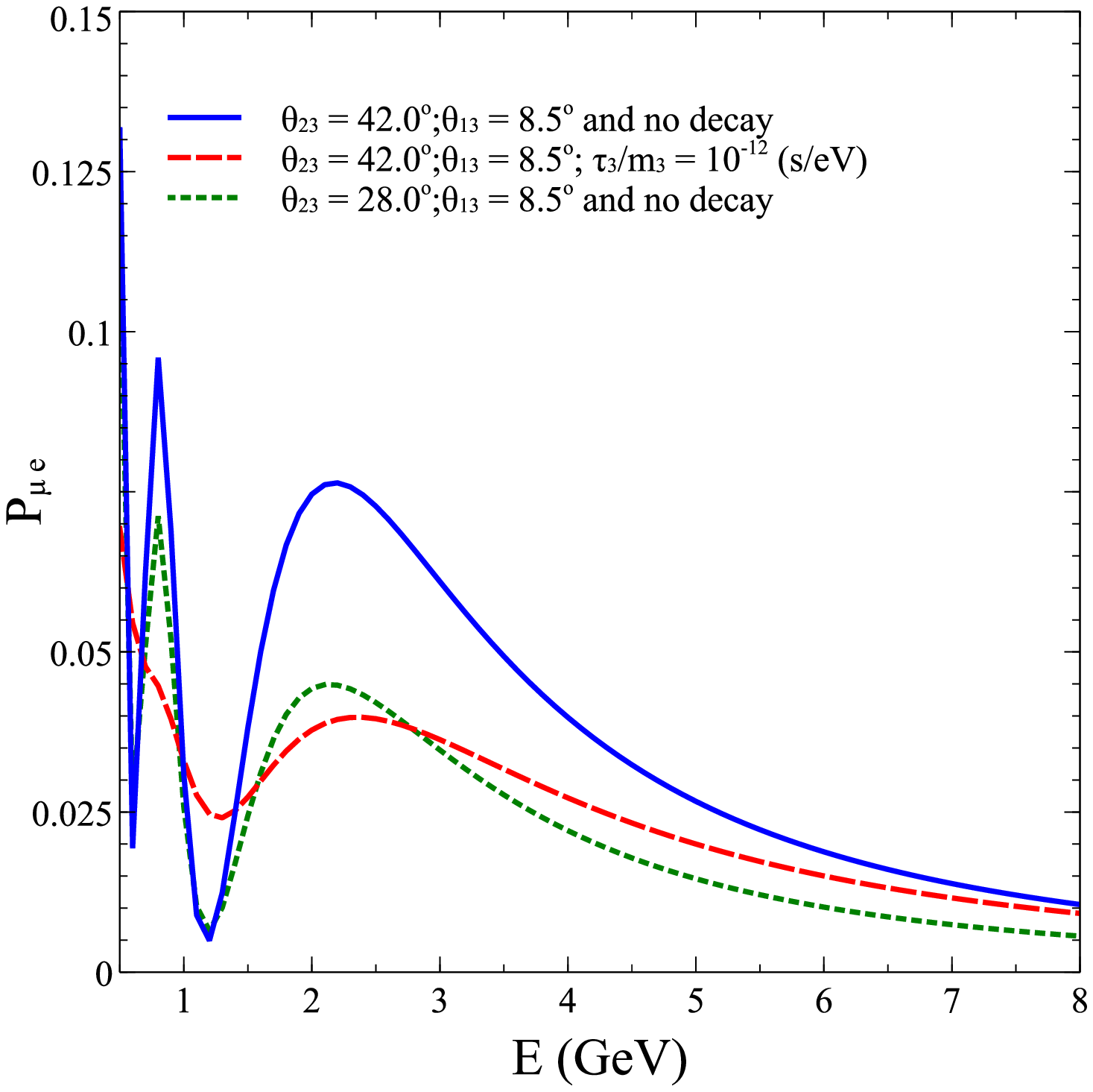}
\includegraphics[width=0.45\textwidth]{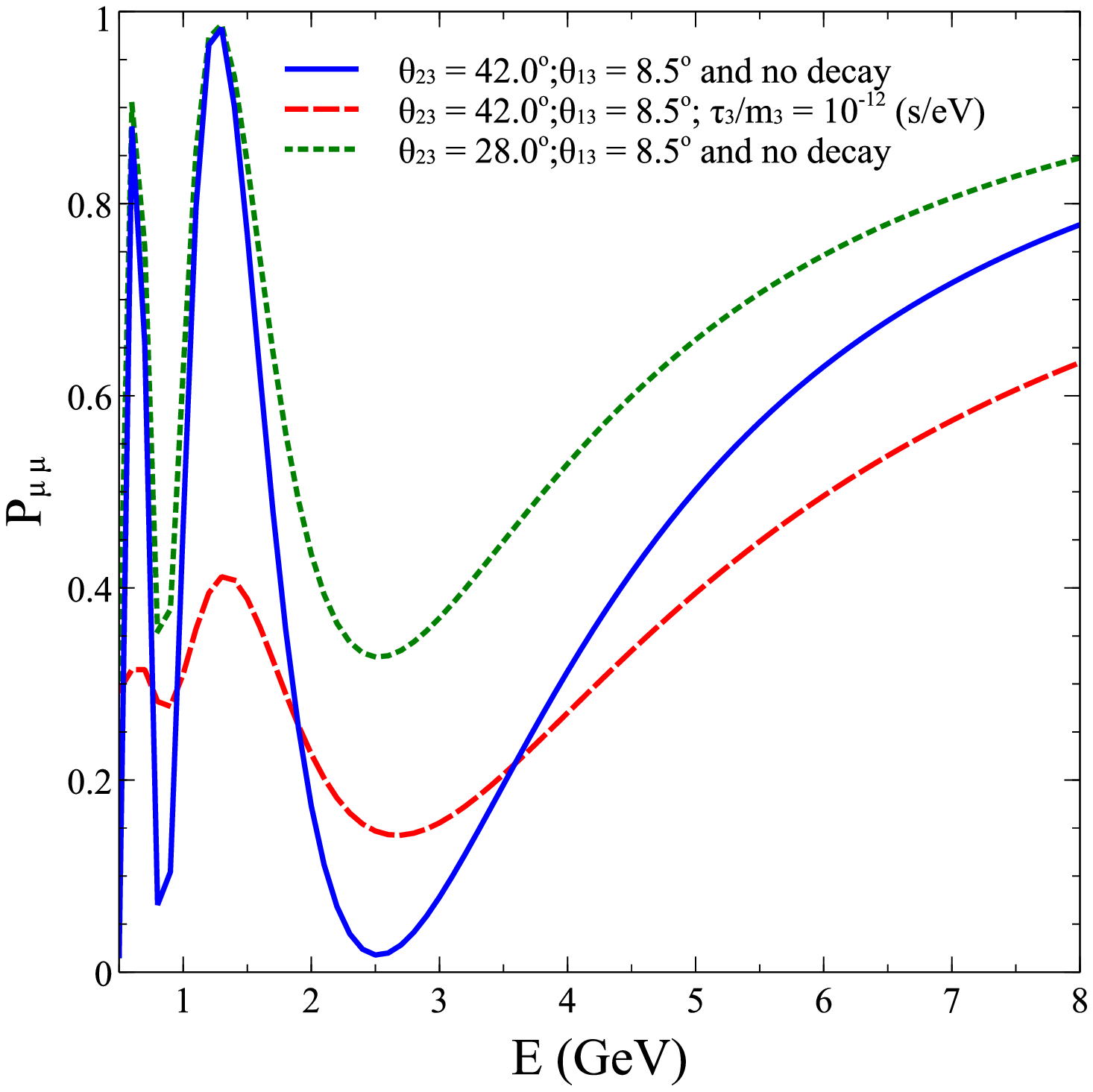}

\caption{\label{fig:prob} The appearance (left panels) and disappearance (right panels) channel neutrino probabilities as a function of neutrino energy. The different lines are described in the legends and also in the text. The top panels show the effect of $\nu_3$ decay for a larger value of $\tau_3/m_3$ while the bottom panels show the effect for a smaller value of $\tau_3/m_3$. }
\end{center}
\end{figure} 

\begin{figure}[]
\begin{center}
\includegraphics[width=0.45\textwidth]{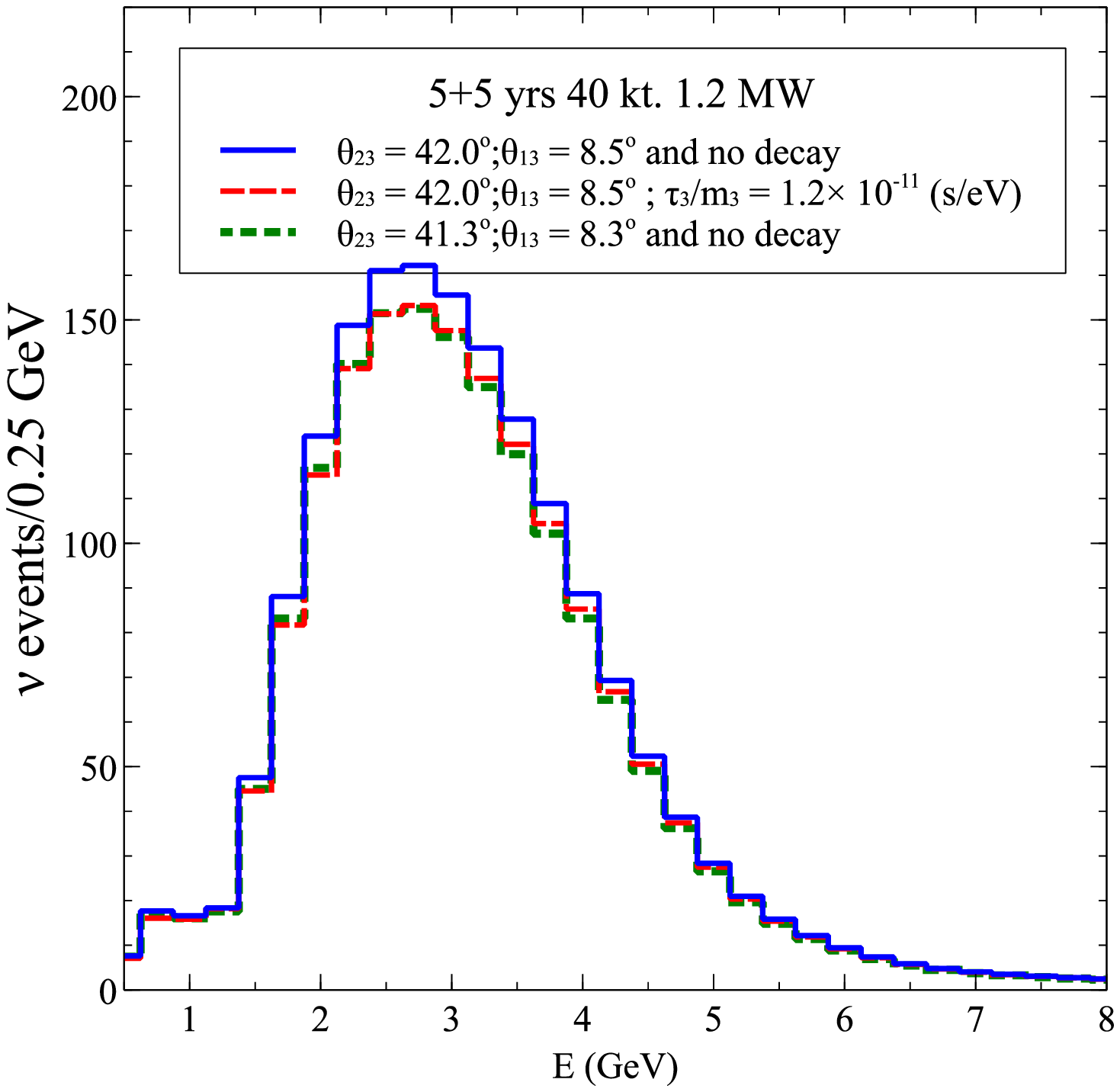}
\includegraphics[width=0.45\textwidth]{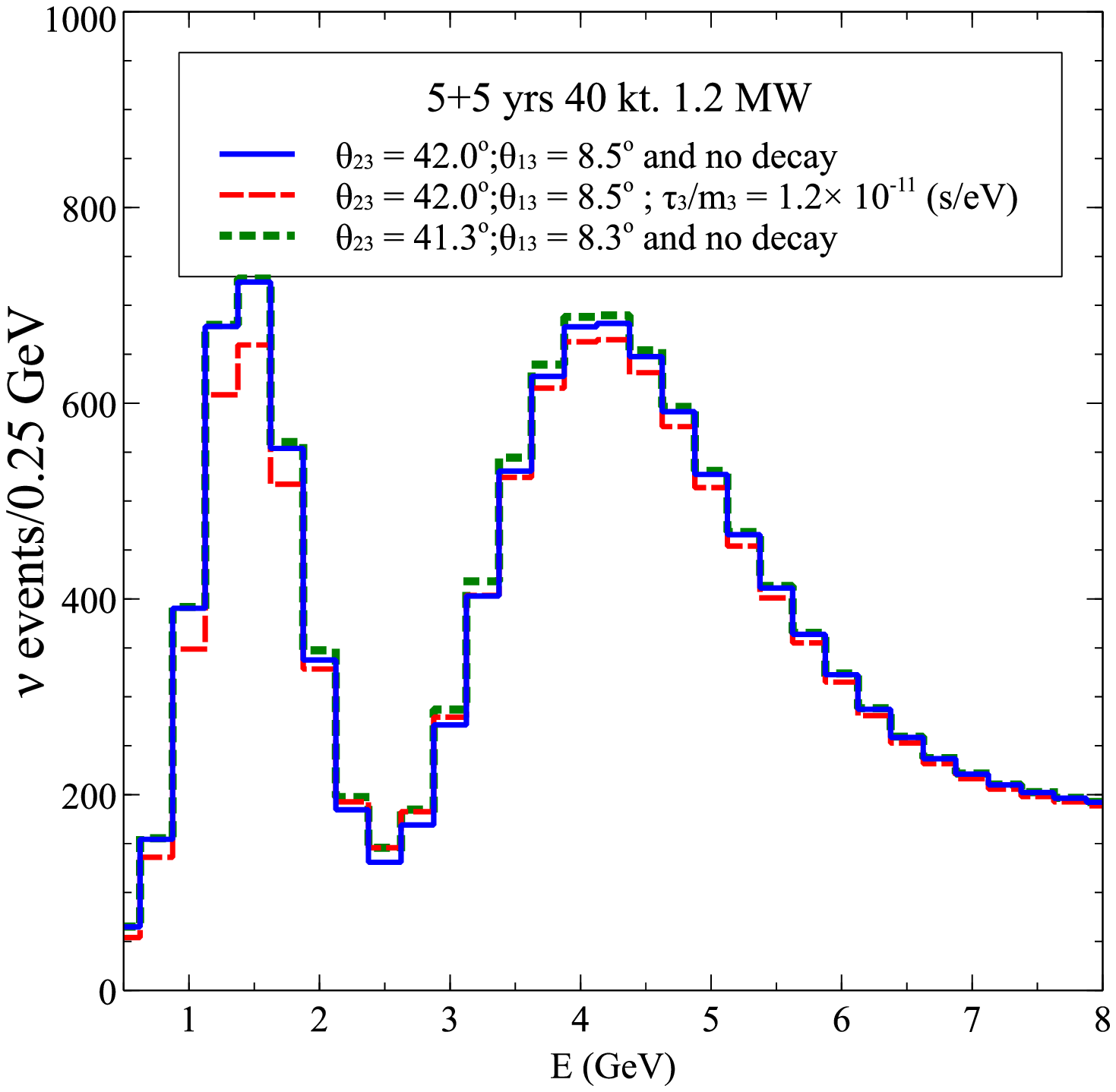}
\includegraphics[width=0.45\textwidth]{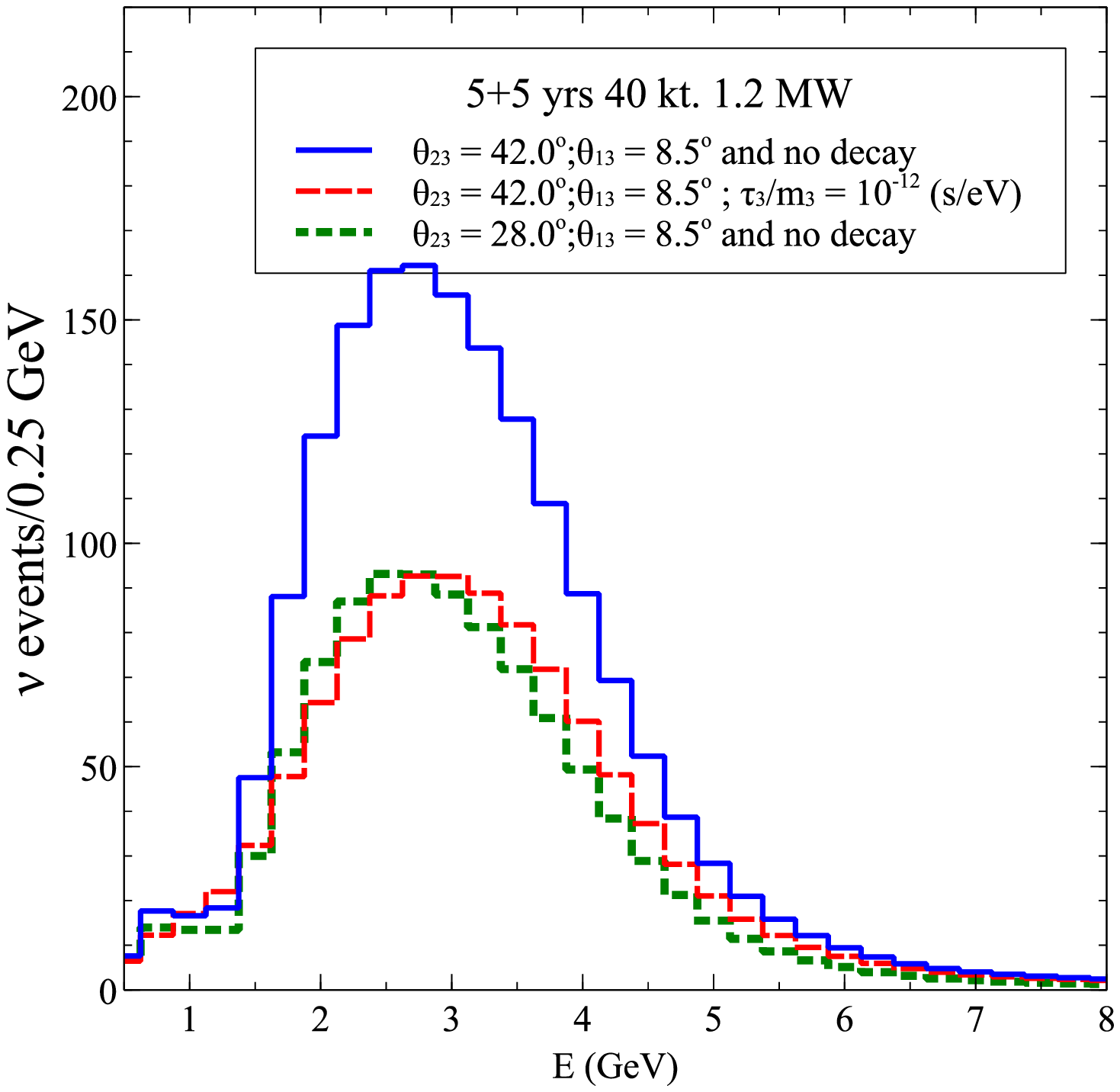}
\includegraphics[width=0.45\textwidth]{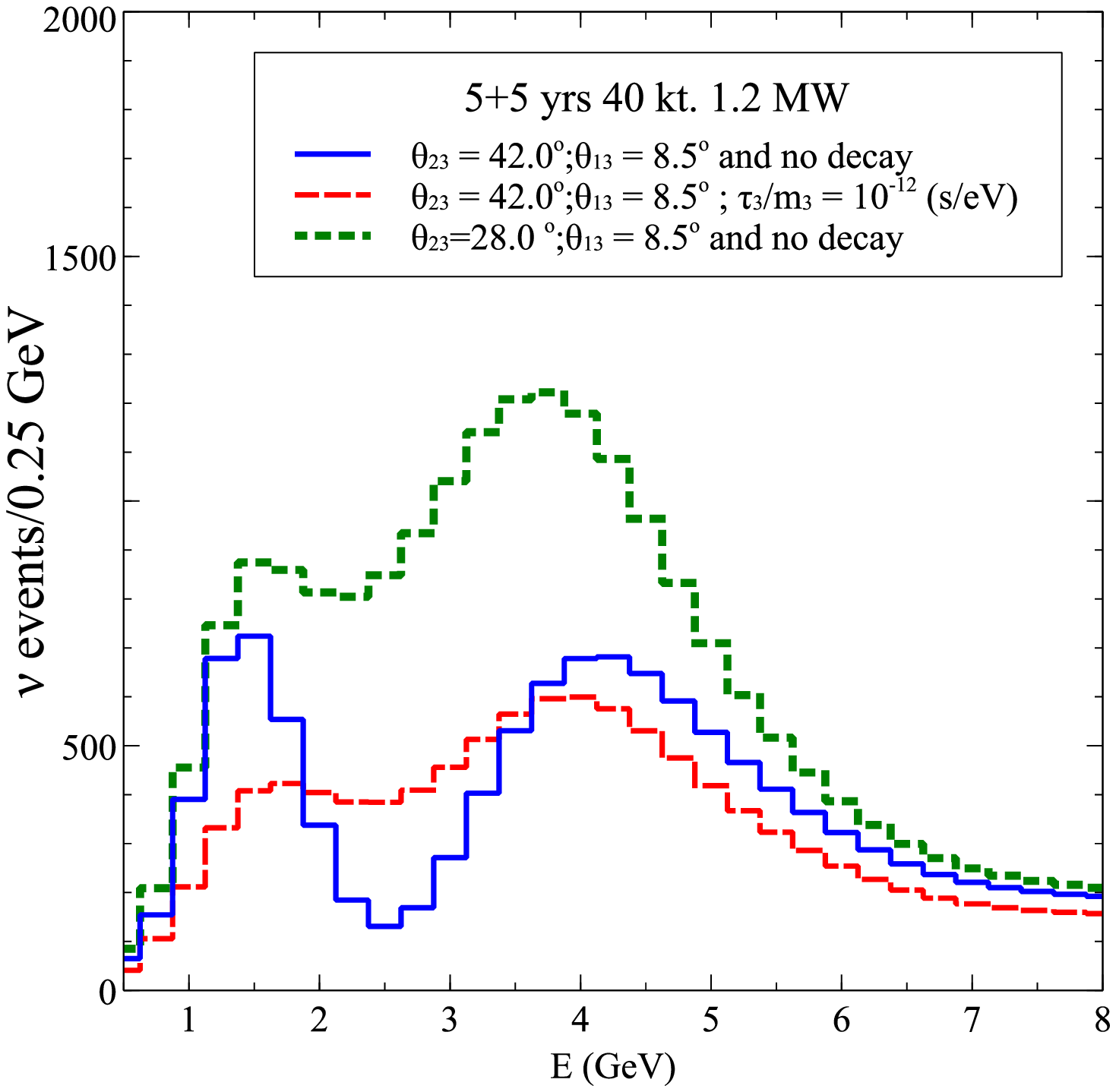}

\caption{\label{fig:event} Same as in Fig.~\ref{fig:prob} but showing neutrino events at DUNE instead of probabilities. The left panels show the electron appearance events, while the right panels show the muon disappearance events. The top panels shows the impact of a larger $\tau_3/m_3$, while the bottom panels are for a smaller $\tau_3/m_3$.}
\end{center}
\end{figure} 

Fig.~\ref{fig:prob} gives the neutrino oscillation probability for both the appearance and disappearance channels. The left panels give the probability for the appearance channel while the right panels show the probability for the disappearance channel. The top panels show the impact of decay on the probabilities for $\tau_3/m_3=1.2\times10^{-11}$ s/eV while the bottom panels show the effect when $\tau_3/m_3=1.0\times10^{-12}$ s/eV. The solid lines and the short-dashed lines show the probabilities for the standard stable case. The blue solid lines in all the four panels are for $\theta_{23}=42^\circ$ and $\theta_{13}=8.5^\circ$ and no decay. The change in the probabilities when decay is switched on 
for the same set of oscillation parameters is shown in all the panels by the red long-dashed line. The first thing to note from this figure is that $P_{\mu e}$ decreases while $P_{\mu\mu}$ increases at the oscillation maximum. The opposite trend is seen for the case when the oscillatory term goes to zero. However, net probability for both appearance as well as disappearance channels decrease in the case of decay. This is because in our model the $\nu_3$ decays to invisible states which do not show-up at the detector. As expected, the extent of decrease of the probability increases as we lower $\tau_3/m_3$ or in other words increase the decay rate, as can be seen by comparing the top panels with the bottom ones. For $\tau_3/m_3=1.0\times10^{-12}$ s/eV we see a drastic change in the probability plots, thereby allowing DUNE to restrict these values of $\tau_3/m_3$, as we had seen in the previous section. 

Next we turn to show the correlation between the decay lifetime $\tau_3/m_3$ and the mixing angles, in particular, the mixing angle $\theta_{23}$. In Fig.~\ref{fig:prob} we show that the appearance channel probability $P_{\mu e}$ for the case with decay (shown by the red long-dashed lines) can be mimicked to a large extent by the no decay scenario if we reduce the value of $\theta_{23}$. These probability curves are shown by the green short-dashed lines. For the top panels we can achieve reasonable matching between the decay and no decay scenario if the $\theta_{23}$ is reduced from $42^\circ$ to $41.3^\circ$ and $\theta_{13}$ is slightly changed from $8.5\degree$ to $8.3\degree$. The matching is such that the green dashed lines is hidden below the red dashed lines in the top panels. In the lower panels, since the $\nu_3$ lifetime is chosen to be significantly smaller, we see a more drastic effect of $\theta_{23}$. In the lower panels the with decay case for $P_{\mu e}$ at $\theta_{23}=42^\circ$ can be somewhat matched by the no decay case if we take a much reduced $\theta_{23}=28^\circ$. However, the disappearance channel is not matched between the red long-dashed and green short-dashed line for the value of $\theta_{23}$ that is needed to match the appearance channel for the decay and no decay cases. 

This correlation between $\tau_3/m_3$ and $\theta_{23}$ in $P_{\mu e}$ can be understood as follows. No decay corresponds to infinite $\tau_3/m_3$. As we reduce $\tau_3/m_3$, $\nu_3$ starts to decay into invisible states reducing the net $P_{\mu e}$ around the oscillation maximum. This reduction increases as we continue to lower $\tau_3/m_3$. On the other hand, it is well known that $P_{\mu e}$ increases linearly with $\sin^2\theta_{23}$ at leading order. Therefore, it is possible to obtain a given value of $P_{\mu e}$ either by reducing $\tau_3/m_3$ or by reducing $\sin^2\theta_{23}$. Therefore, it will be possible to compensate the decrease in $P_{\mu e}$ due to decay by increasing the value of $\sin^2\theta_{23}$. Hence, if we generate the appearance data taking decay, we will be able to fit it with a theory for stable neutrinos by suitably reducing the value of $\sin^2\theta_{23}$.

The correlation between $\tau_3/m_3$ and $\theta_{23}$ for the survival channel on the other hand is complicated. For simplicity, let us understand that within the two-generation framework first, neglecting matter effects. The effect of three-generations will be discussed a little later and the effect of earth matter is not crucial for the DUNE energies in this discussion. The survival probability in the two-generation approximation is given by \cite{Gomes:2014yua}
\be
P_{\mu\mu}^{2G} = \left [ \cos^2\theta_{23} + \sin^2\theta_{23} e^{-\frac{m_3 L}{\tau_3 E}} \right ]^2 - 
\sin^22\theta_{23} e^{-\frac{m_3 L}{2\tau_3 E}}\,\sin^2\left(\frac{\Delta m_{31}^2 L}{4E}\right)
\,.
\label{eq:pmm}
\ee
The Eq.~(\ref{eq:pmm}) shows that decay affects both the oscillatory term as well as the constant term in  $P_{\mu\mu}$, causing both to reduce. Therefore, it is not difficult to see that with decay included, the value of $\theta_{23}$ should be increased to get the same $P_{\mu\mu}$ as in the no decay case. Hence, in this case again if we generate the disappearance data taking decay, we will be able to fit it with a theory for stable neutrinos by suitably reducing the value of $\theta_{23}$. However, note that the dependence of $P_{\mu\mu}$ on $\theta_{23}$ and $\tau_3/m_3$ is different from the dependence of $P_{\mu e}$ on $\theta_{23}$ and $\tau_3/m_3$ and hence we never get the same fitted value of $\theta_{23}$ for the two channels. This is evident in Fig.~\ref{fig:prob} where in the lower panel the appearance probability fits between decay case and $\theta_{23}=42^\circ$ and no decay case and $\theta_{23}=28^\circ$. However, this does not fit the disappearance probability simultaneously.  One can check that the above understanding of the correlation between $\theta_{23}$ and $\tau_3/m_3$ is true for the full three-generation case too. We will show below the probabilities for the full three-generation case with decay and matter effects obtained by an exact numerical computation.

In order to see the correlation between $\tau_3/m_3$ and $\theta_{23}$ at the event level, we plot in Fig.~\ref{fig:event} the appearance (electron) and disappearance (muon) events for 5 years of running of DUNE in the neutrino channel. One will expect a similar behaviour for the antineutrinos as well. The respective panels and the three plots in each panel are arranged in exactly the same way as in Fig.~\ref{fig:prob}. We note that all the features that were visible at the probability level in Fig.~\ref{fig:prob} are also seen clearly at the events level in Fig.~\ref{fig:event}. Neutrino events are seen to reduce with the onset of neutrino decay, with the extent of reduction increasing sharply with the value of the decay rate ($1/\tau_3$). The electron event spectrum for the case with decay can be seen to be roughly mimicked with that without decay but with a lower value of the $\theta_{23}$, the required change in the value of $\theta_{23}$ increasing with the decay rate ($1/\tau_3$). 
On the other hand for the lower panel, the muon spectrum in absence of decay (shown by the green lines) would need a different value of $\theta_{23}$ to match the muon spectrum in presence of decay (shown by the red lines). This mismatch between the fitted value of $\theta_{23}$ between the appearance and disappearance channels can hence be expected to be instrumental in breaking the approximate degeneracy between $\tau_3/m_3$ and $\theta_{23}$.

\begin{figure}[h]
\begin{center}
\includegraphics[width=0.32\textwidth]{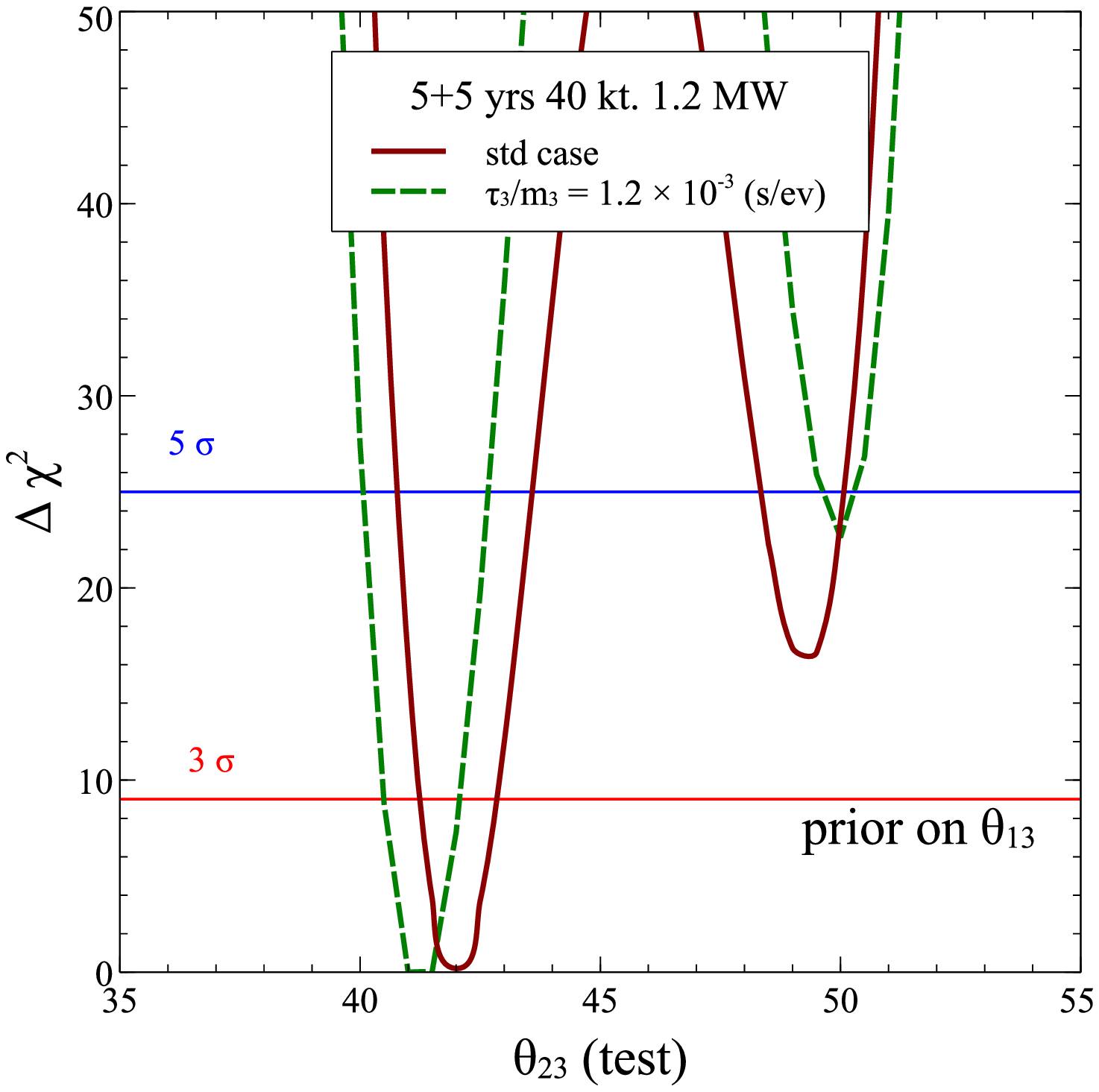}
\includegraphics[width=0.32\textwidth]{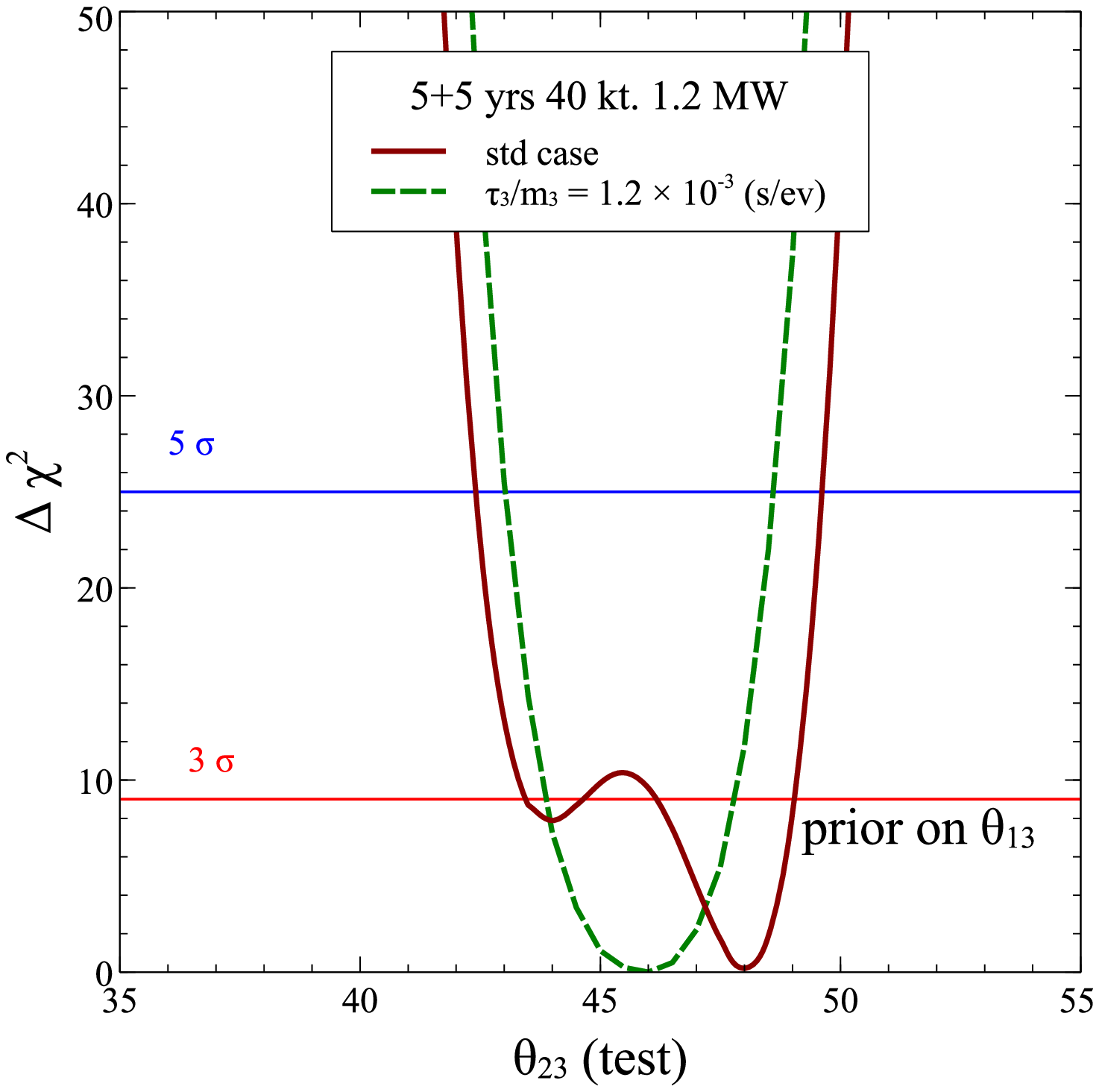}
\includegraphics[width=0.32\textwidth]{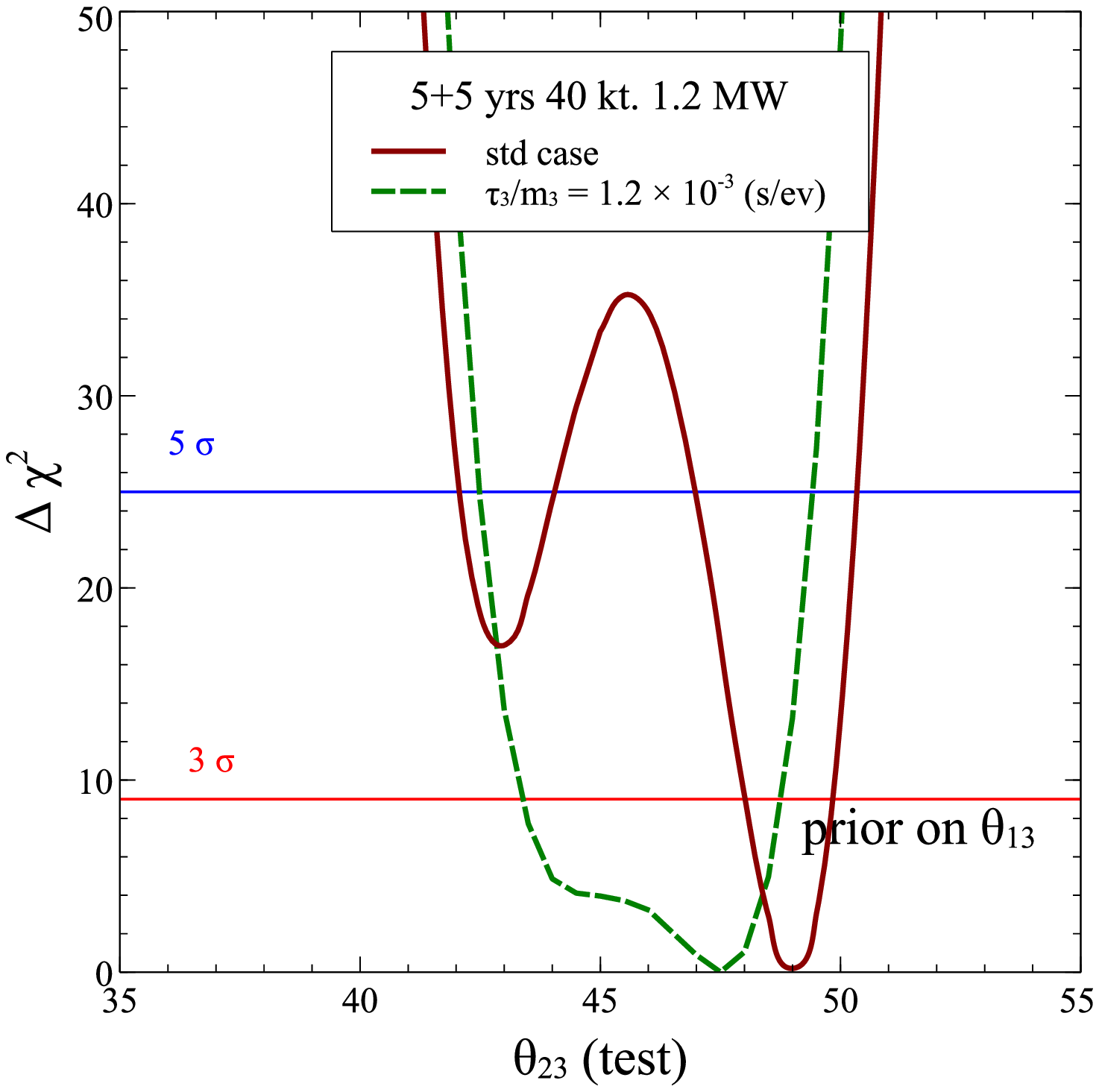}
\caption{\label{fig:th23}$\chi^2$ as a function of  $\theta_{23}$(test). The left, middle and right panels are for the cases when the data is generated at $\theta_{23}=42^\circ$, $\theta_{23}=48^\circ$ and $\theta_{23}=49.3^\circ$, respectively. The dark red solid curves are for the standard case when both data and fit are done within the three-generation framework of stable neutrinos. The green dashed curves are for the case when the data is generated for unstable $\nu_3$ with $\tau_3/m_3=1.2\times 10^{-11}$ s/eV but it is fitted assuming stable neutrinos. }
\end{center}
\end{figure} 

\begin{figure}[h]
\begin{center}
\includegraphics[width=0.45\textwidth]{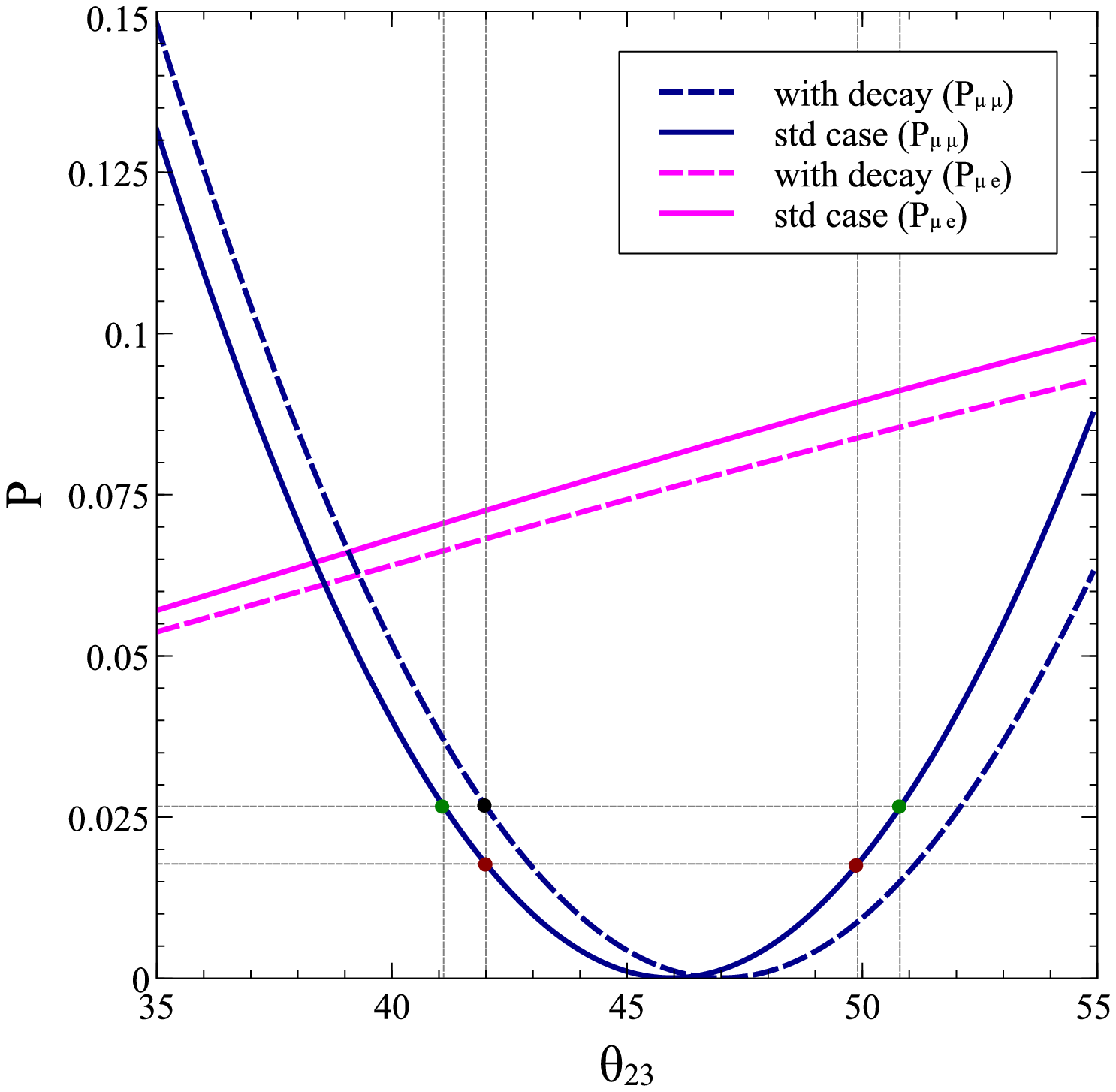}
\includegraphics[width=0.45\textwidth]{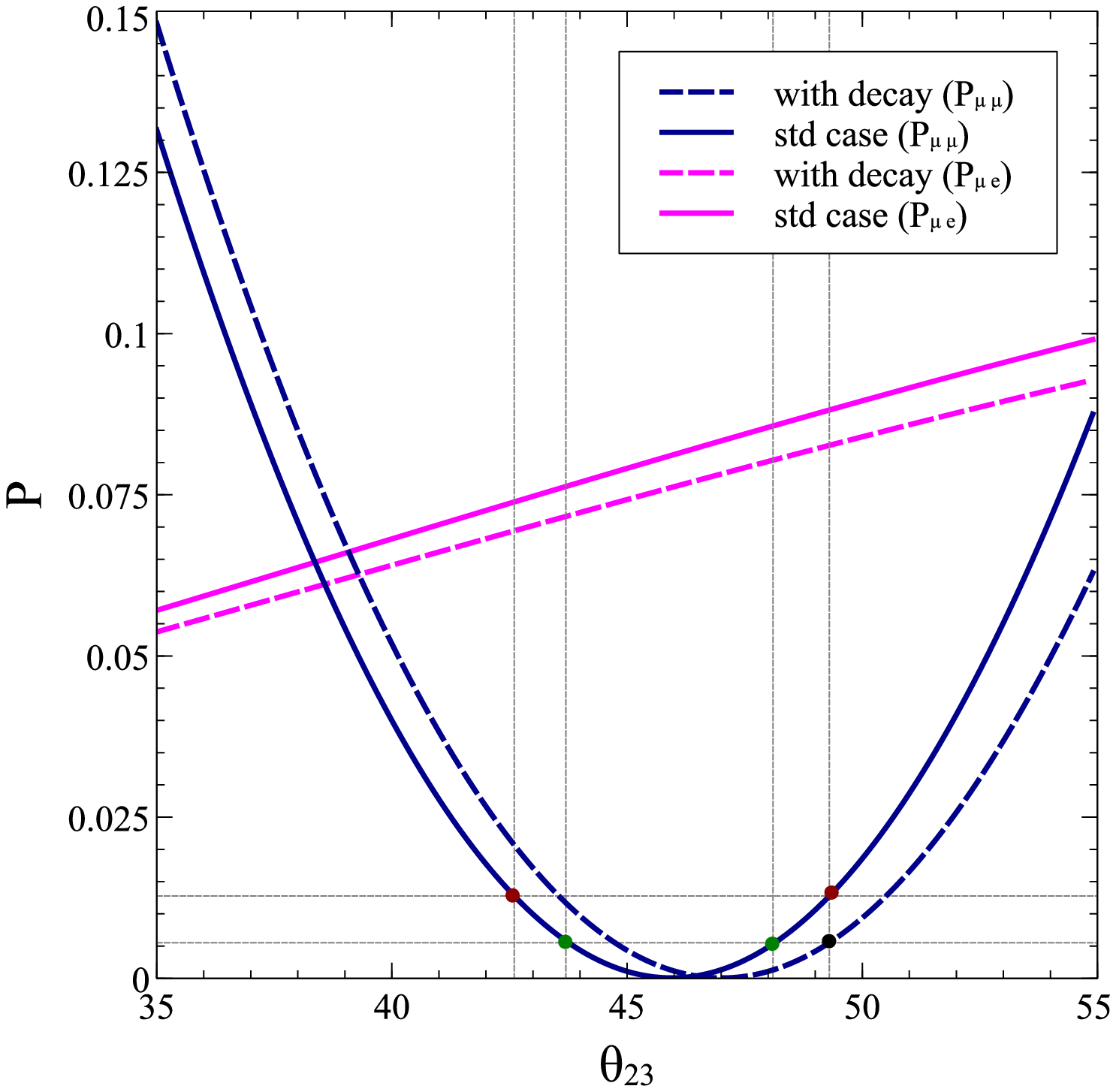}
\caption{\label{fig:probloho}The probabilities $P_{\mu\mu}$ (blue lines) and $P_{\mu e}$ (magenta lines), shown as a function of  $\theta_{23}$. The plots have been drawn for the DUNE baseline and $E=2.5$ GeV, taking all oscillation parameters as mentioned in section III. The left and right panels are identical apart from the horizontal and vertical lines which show the probabilities for the cases when the data is generated at $\theta_{23}=42^\circ$ (left panel) and $\theta_{23}=49.3^\circ$ (right panel). The solid curves show the probabilities for the standard case while the dashed curves are for the case for unstable $\nu_3$ with $\tau_3/m_3=1.2\times 10^{-11}$ s/eV. The probabilities are shown for the full three-generation framework including earth matter effect. The black dot shows the point at which data is generated for the decay case, while the green dots show the points which give the same $P_{\mu\mu}$ as the black dot, but for the standard case. The red dots show the $P_{\mu\mu}$ for the data generated for the standard case at $\theta_{23}=42^\circ$ (left panel) and $\theta_{23}=49.3^\circ$ (right panel) and the corresponding fake minima when fitted by the standard case.}
\end{center}
\end{figure} 

In order to study the impact of decay on the expected $\theta_{23}$ sensitivity of DUNE, we show in Fig.~\ref{fig:th23} the $\chi^2$ as a function of $\theta_{23}$(test). The left panel is for the case when the data is generated at $\theta_{23}=42^\circ$, middle panel is for data at $\theta_{23}=48^\circ$ and the right panel is for data at $\theta_{23}=49.3^\circ$. The dark red solid curves are for the standard case when both data and fit are done within the three-generation framework of stable neutrinos. The green dashed curves are for the case when the data is generated for unstable $\nu_3$ with $\tau_3/m_3=1.2\times 10^{-11}$ s/eV but it is fitted assuming stable neutrinos. For generating the data all other oscillation parameters are taken as mentioned before in section III. The fits are marginalised over $\theta_{13}$, $\dcp$ and $\Delta m^2_{31}$ in their current $3\sigma$ ranges. Before we proceed to look at the impact of decay on the measurement of $\theta_{23}$ at DUNE, let us expound some features of $\theta_{23}$ for standard three-generation oscillation scenario. A comparison of the red curves in the three panels of  Fig.~\ref{fig:th23} shows that the left panel and the right panel look like near mirror images of each other, while the middle panel looks different. Note that $42^\circ$ is as far removed from $\theta_{23}=45^\circ$ as $48^\circ$, however the $\chi^2$ curves for the $\theta_{23}=42^\circ$ and $\theta_{23}=48^\circ$ cases appear different. This is due to three-generation effects coming from the non-zero $\theta_{13}$ \cite{Raut:2012dm}.  The values of $\theta_{23}$ in HO and LO that correspond to the same effective mixing angle $\theta_{\mu\mu}$ and which gives the same $P_{\mu\mu}$ are given as \cite{Raut:2012dm}
\be
\sin\theta_{23}^{LO} = \frac{\sin\theta_{\mu\mu}^{LO}}{\cos\theta_{13}}\,\,\,\,&;&\,\,\,\,\sin\theta_{23}^{HO} = \frac{\sin\theta_{\mu\mu}^{HO}}{\cos\theta_{13}} \label{eq:th231} \\
\theta_{\mu\mu}^{LO} &=& 90^\circ - \theta_{\mu\mu}^{HO}
\,,
\label{eq:th23}
\ee
which gives $\theta_{23}=49.3^\circ$ as the mixing angle that gives the same $P_{\mu\mu}$ as $\theta_{23}=42^\circ$ instead of $\theta_{23}=48^\circ$, as we would expect in the two-generation case. 
In order to further illustrate this point, we show in Fig.~\ref{fig:probloho} the survival probability $P_{\mu\mu}$ (blue lines) as a function of $\theta_{23}$ for the standard case (solid line) and decay case (dashed line). Also shown are the corresponding oscillation probability $P_{\mu e}$ (magenta lines) for the standard case (solid line) and decay case (dashed line). The plots have been drawn for the DUNE baseline and $E=2.5$ GeV, taking all oscillation parameters as mentioned in section III. The energy 2.5 GeV corresponds to oscillation maximum at the DUNE baseline where the DUNE flux peaks. We note that for the standard oscillations case,  $P_{\mu\mu} \simeq 0$ corresponds to a value of $\theta_{23} \simeq 46^\circ$ and not $45^\circ$ as in the two-generation case.
We also note that $P_{\mu\mu}$ at $\theta_{23}=42^\circ$ in LO is matched by the $P_{\mu\mu}$ at $\theta_{23}\simeq49.9^\circ$ in HO, the small difference between the value of $\theta_{23}^{HO}$ derived from Eq.~(\ref{eq:th23}) and the exact numerical results shown in  Fig.~\ref{fig:probloho} come from earth matter effects mainly. 

The solid red curves in Fig.~\ref{fig:th23} showing the $\chi^2$ vs. $\theta_{23}$(test) for the standard oscillation case match well with the solid blue probability curves in Fig.~\ref{fig:probloho}. For the left panel, data is generated at $\theta_{23}=42^\circ$ and the absolute and fake minima come at $\theta_{23}({\rm test})=42^\circ$ and $49.5^\circ$, respectively. On the other hand for the right panel, data is generated at $\theta_{23}=49.3^\circ$ and the absolute and fake minima come at $\theta_{23}({\rm test})=49.3^\circ$ and $43.0^\circ$, respectively. Note that since $P_{\mu\mu}$  is nearly matched at the true and fake minima points, the disappearance data would return a $\chi^2 \simeq 0$ at both the true as well as fake minima points giving an exact octant degeneracy. The main role of the disappearance data is only to determine the position of the minima points in $\theta_{23}$. The oscillation probability $P_{\mu e}$ on the other hand is very different between the true and fake minima points as can be seen from the solid magenta line in Fig.~\ref{fig:probloho}. Hence, the appearance channel distinguishes between the two and gives a non-zero $\chi^2$ at the fake minima and breaks the octant degeneracy. We can see from Fig.~\ref{fig:th23} that for the $\theta_{23}=42^\circ$ case (left panel), the $\chi^2$ corresponding to the wrong octant minima is 16.6 while for $\theta_{23}=49.3^\circ$ case (right panel) it is 17.1. Hence, for standard oscillation the octant sensitivity at $\theta_{23}=42^\circ$ is only slightly worse than the octant sensitivity at $\theta_{23}=49.3^\circ$. The reason for this is that the $\chi^2$ for octant sensitivity is given in terms of the difference between the appearance channel event spectra for the true and fake $\theta_{23}$ points. 
One can see from the solid magenta lines in Fig.~\ref{fig:probloho} that this difference is almost the same for the left and right panels for the standard oscillations case and hence 
the $\chi^2$ of the fake minima for the solid red lines in the left and right panels in Fig.~\ref{fig:th23} are nearly the same. The octant sensitivity for the middle panel ($\theta_{23}=48^\circ$) is significantly poorer since for this case, the difference in the appearance channel probability is much smaller. This happens because this value of $\theta_{23}$ is too close to effective maximal mixing for $P_{\mu\mu}$ (cf. Eqs.~(\ref{eq:th231} and (\ref{eq:th23})).

Next we look at the impact of including neutrino decay in data on $\theta_{23}$ measurement at DUNE, shown by the green dashed lines in Fig.~\ref{fig:th23}. These lines are obtained by generating data including decay but fitting them with standard three-generation oscillations with stable neutrinos. We notice that compared to the red solid lines for the standard case, the position of minima as well as the $\chi^2$ at the fake minima have changed. For the left panel ($\theta_{23}=42^\circ$ in data) the minima points shift to $\theta_{23}({\rm test})=41.0^\circ$ in LO and $\theta_{23}({\rm test})=50.0^\circ$ in HO. Thus, for data with $\theta_{23}$ in LO, the minima point shifts to lower $\theta_{23}$(test) in LO and higher $\theta_{23}$(test) in HO. On the other hand for right panel ($\theta_{23}=49.3^\circ$ in data) the minima points shift to $\theta_{23}({\rm test})=44.5^\circ$ in LO and $\theta_{23}({\rm test})=47.5^\circ$ in HO. Thus, for data with $\theta_{23}$ in HO, the minima point shifts to higher $\theta_{23}$(test) in LO and lower $\theta_{23}$(test) in HO. Note that none of the minima now correspond to the true value of $\theta_{23}$ at which the data is generated. Note also that the gap between the two minima points has increased for the case with data in LO and decreased for the case with data in HO.

The shifting of minima for both the LO and HO data points can be understood easily in terms of the left and right panels of Fig.~\ref{fig:probloho}, respectively. This figure shows $P_{\mu\mu}$ (and $P_{\mu e}$) at the oscillation maximum as a function of $\theta_{23}$. The solid lines are for no decay while the dashed lines are for decay and oscillations. An important thing we can note in this figure is that with decay the $P_{\mu\mu}$ curve gets shifted towards the right. Even the effective maximal mixing point gets shifted further towards higher values of $\theta_{23}$. The left panel of Fig.~\ref{fig:probloho} shows the data point for the disappearance channel for $\theta_{23}=42^\circ$ by the black point on the blue dashed line, which includes decay. This $P_{\mu\mu}$ has to be reproduced by the no decay theory in the fit. The corresponding minima points can be obtained by following the blue solid line and are shown by the green dots, at $\theta_{23}=41.1^\circ$ and $\theta_{23}=50.8^\circ$. This matches well with the minima on the green dashed line in the left panel of Fig.~\ref{fig:th23}. The point where data is generated for the HO case in Fig.~\ref{fig:th23} is shown by the black dot in the right panel of Fig.~\ref{fig:probloho}. The corresponding fit points coming from $P_{\mu\mu}$ can be seen at the green dots in this panel at $\theta_{23}=43.7^\circ$ and $\theta_{23}=48.1^\circ$. Note that since the decay causes the $P_{\mu\mu}$ curve to shift towards the right, all minima points in $\theta_{23}$ are shifted towards the left. However, when one compares the minima in the true and fake octants, it turns out that the two minima get further separated for the LO case (left panel), while for the HO case (right panel) they come closer together. This is consistent with the gap between the minima increasing for the left panel and decreasing for the right panel in Fig.~\ref{fig:th23}. As mentioned before, the disappearance data plays the role of determining the minima in the true and fake octant, but brings no significant octant sensitivity since $P_{\mu\mu}$ can be  matched at the true and fake minima, at least at the oscillation maximum. The octant sensitivity comes from the difference in the number of appearance events at the minima points at the true and fake octant. Therefore, since the minima points in the LO case gets further separated for the decay case compared to no decay, the octant sensitivity for LO increases as can be seen from the $P_{\mu e}$ curve in Fig.~\ref{fig:probloho}. We can read from the left panel of  Fig.~\ref{fig:th23} that the $\chi^2$ at the minima in the fake octant is 22.6, higher than the case for standard oscillations. On the other hand for the right panel, the minima points come closer in the case of the dashed green lines and the octant sensitivity coming from the appearance channel drops significantly to $\chi^2=4.1$ for the wrong octant since the difference in $P_{\mu e}$ between the minima points reduces, as can be seen from the right panel of  Fig.~\ref{fig:probloho}.

\begin{figure}[t!]
\begin{center}
\includegraphics[width=0.32\textwidth]{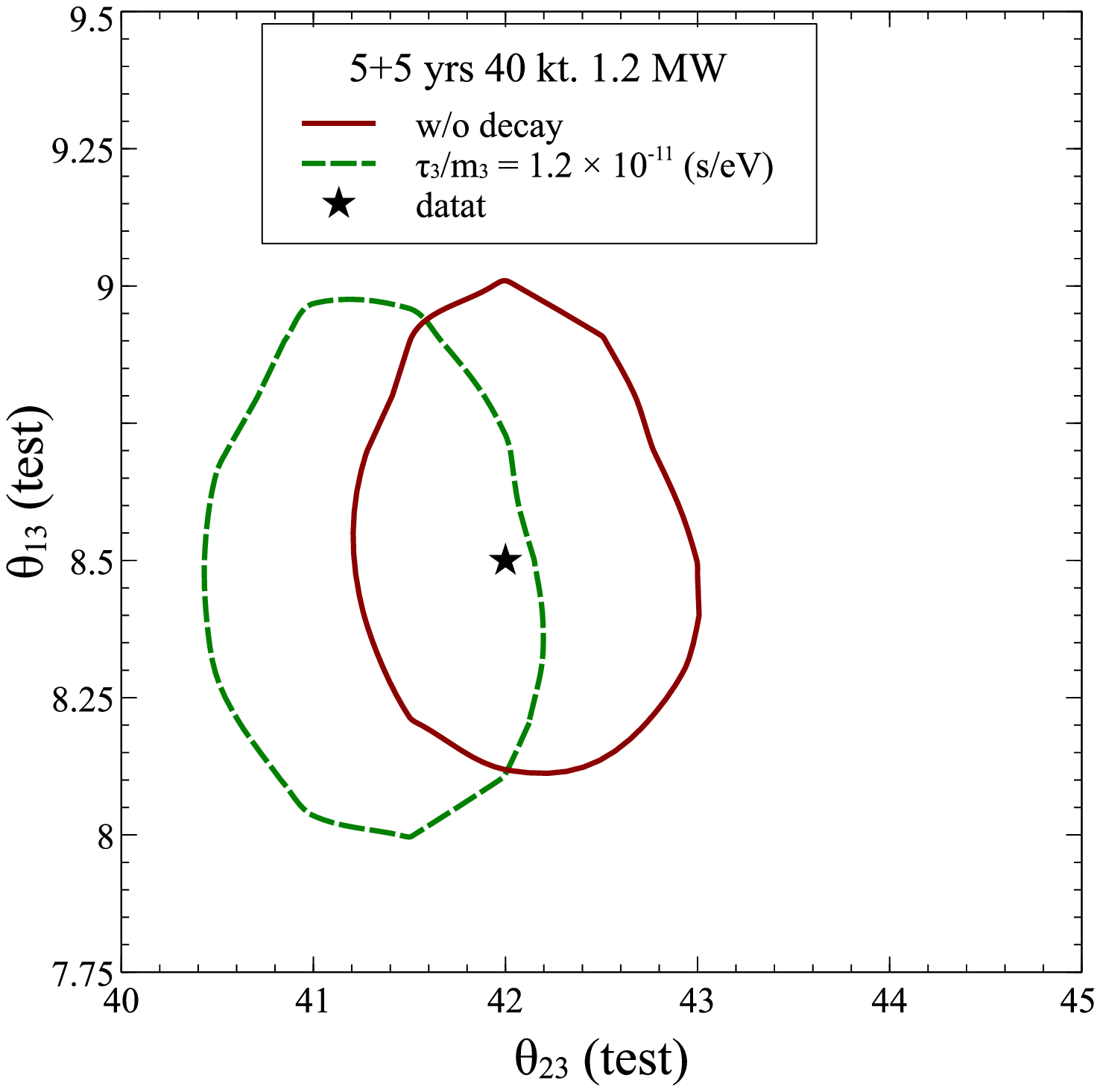}
\includegraphics[width=0.32\textwidth]{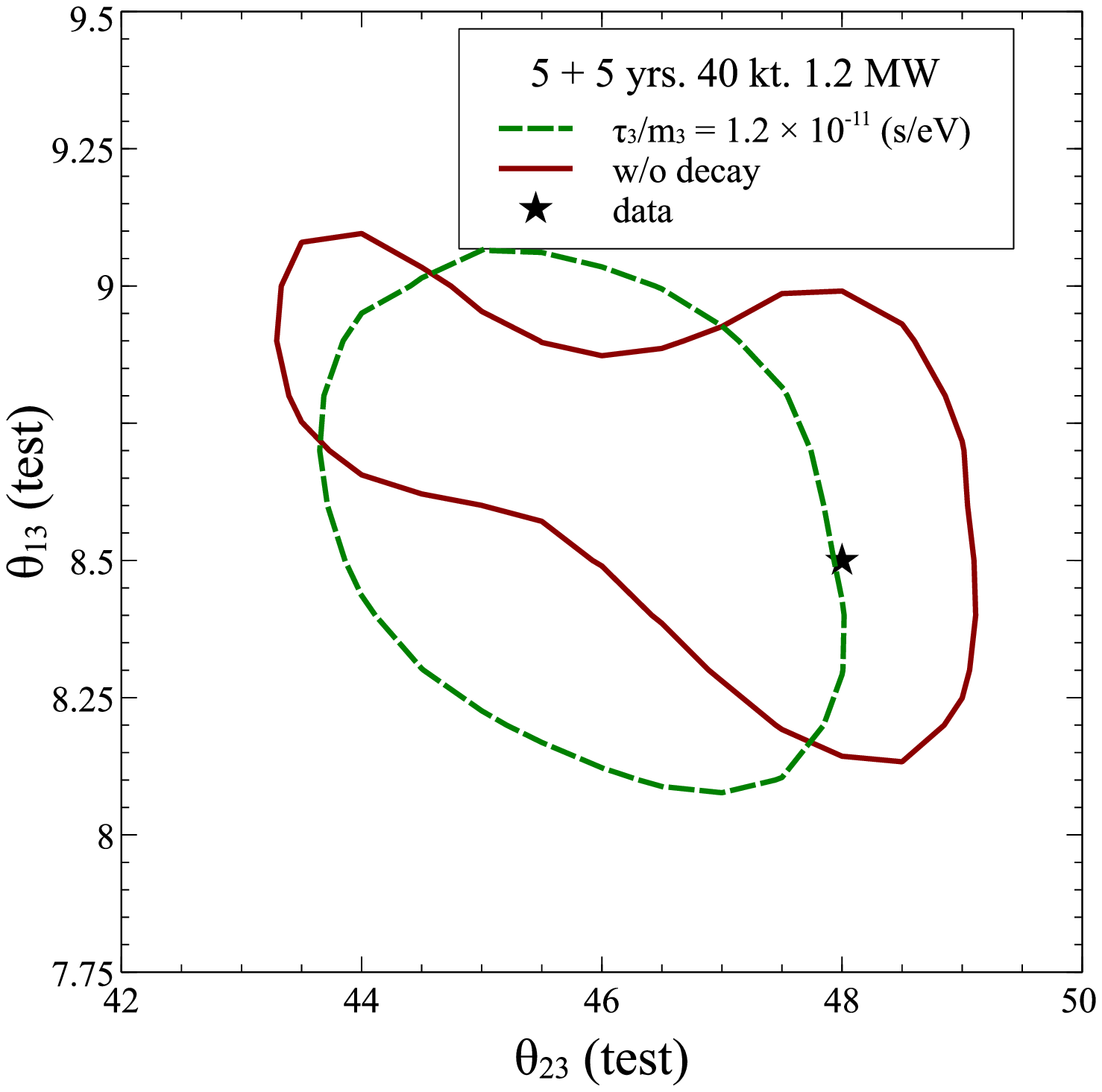}
\includegraphics[width=0.32\textwidth]{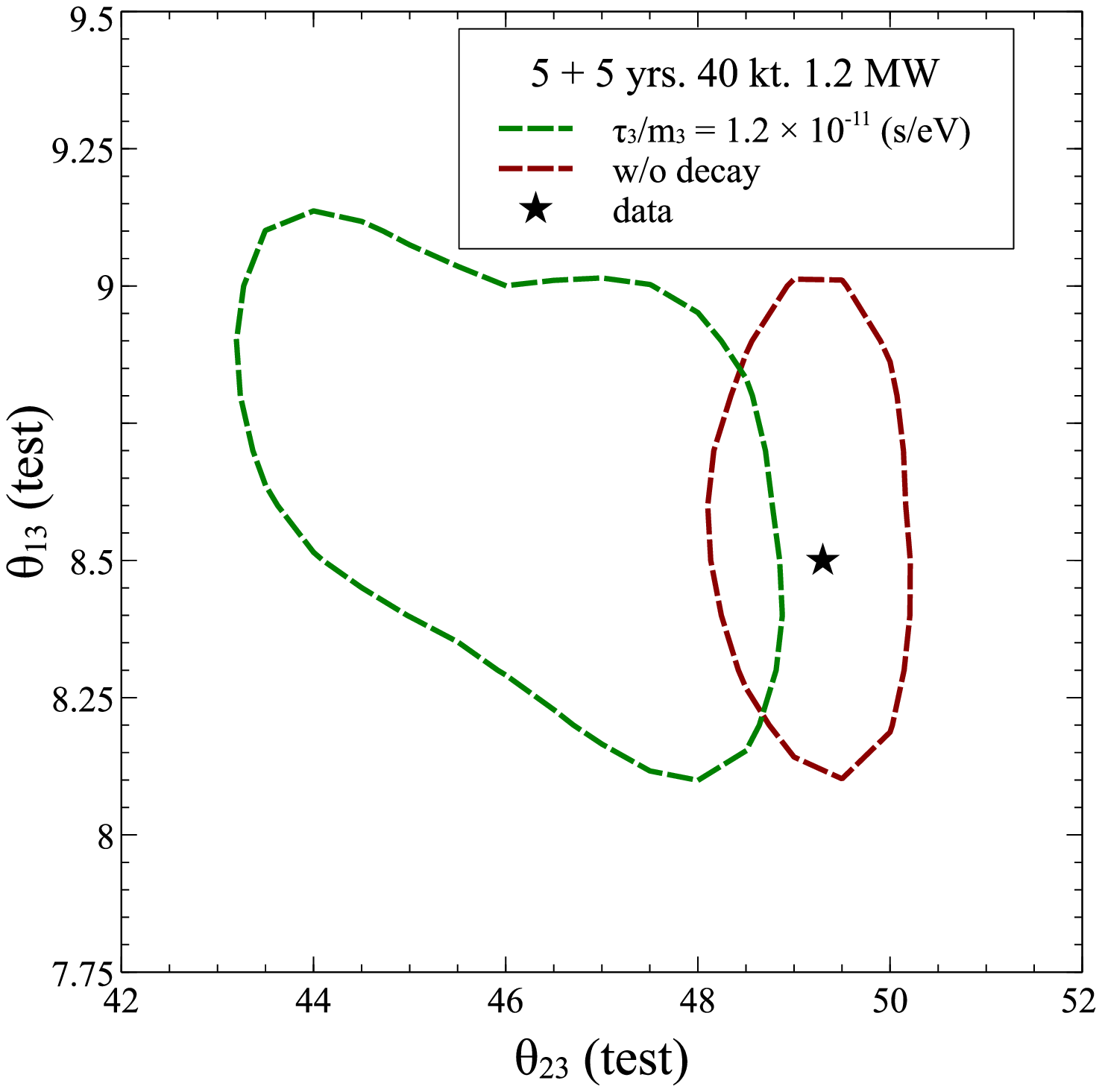}
\caption{\label{fig:th23th13}The plots show the expected $3\sigma$ C.L. contours in 
the $\theta_{23}-\theta_{13}$ plane
for the case when the data is simulated at $\theta_{23}=42\degree$ (left panel), $\theta_{23}=48\degree$ (middle panel), and $\theta_{23}=49.3\degree$ (right panel). The value of $\theta_{13}=8.5^\circ$ in all panels. The black stars show the data points in the plane. The dark red solid curves show the expected 3$\sigma$ contour for the standard scenario in absence of decay in data and theory. The green dashed curves show the 3$\sigma$ contour for the case when the data corresponds to a decaying $\nu_3$ with $\tau_{3}/m_{3}=1.2\times10^{-11}$ s/eV, which is fitted with a theory where all neutrinos are taken as stable. }
\end{center}
\end{figure} 

The impact of decay on the expected constraints in the two-dimensional $\theta_{23}-\theta_{13}$ plane is shown in Fig.~\ref{fig:th23th13}. As in Fig.~\ref{fig:th23}, the left panel shows the results when the data is generated at $\theta_{23}=42^\circ$, middle panel is for $\theta_{23}=48^\circ$, while the right panel gives the results for data corresponding to $\theta_{23}=49.3^\circ$. The point where the data is generated is marked by a star in the $\theta_{23}-\theta_{13}$ plane. The expected contours correspond to 3$\sigma$ C.L. The dark red solid lines are obtained for the standard case when neutrinos are taken as stable in both the data as well as the fit. The green dashed ones are obtained when we simulate the data assuming an unstable $\nu_3$ with $\tau_3/m_3=1.2\times 10^{-11}$ s/eV, but fit it with the standard case assuming stable neutrinos. The contours are marginalized over test values of $\dcp$ and $\Delta m^2_{31}$ within their current $3 \sigma$ ranges. The impact of decay is visible in all panels. Though the contours change in both mixing angles, the impact on $\theta_{23}$(test) is seen to be higher than the impact on $\theta_{13}$(test). As we had seen in details above in Fig.~\ref{fig:th23}, the green contours are shifted to lower values of $\theta_{23}$ in both the left and right panels. The one-to-one correspondence between the allowed $\theta_{23}$(test) values at $3\sigma$ between this figure and Fig.~\ref{fig:th23} can be seen. The mild anti-correlation between the allowed values of $\theta_{23}$(test) and $\theta_{13}$(test) for the green dashed lines comes mainly from the appearance channel which depends on the product of $\sin^2\theta_{23}\sin^22\theta_{13}$ at leading order. This anti-correlation is seen to be more pronounced for the middle and right panels because for these cases the $\theta_{23}$ sensitivity of the data falls considerably in presence of decay and the $\chi^2$ drops.

\begin{figure}
\includegraphics[scale=0.6]{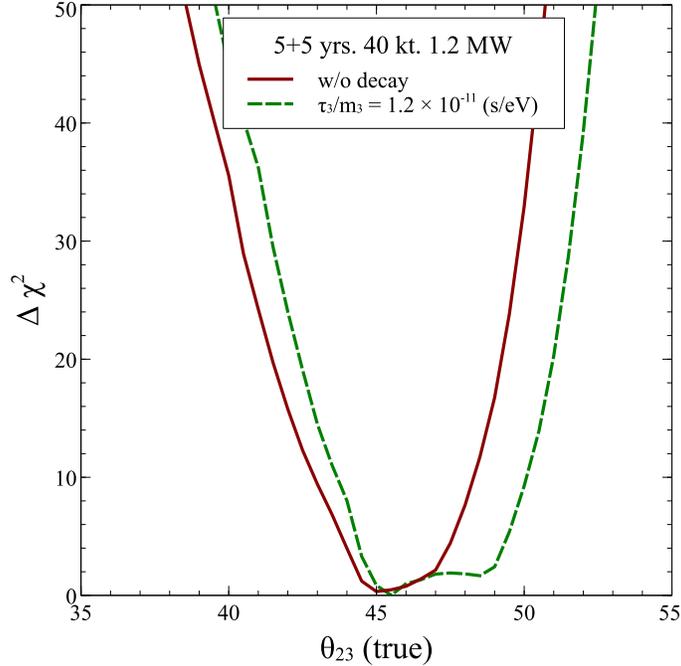}
\caption{\label{fig:octant} Expected octant sensitivity at DUNE. The dark red solid curve is for standard case of stable neutrinos. The green dashed curve is for the case when $\nu_3$ is taken as unstable with $\tau_3/m_3=1.2\times 10^{-11}$ s/eV in the data, but in the fit we keep it to be stable.}
\end{figure}

The Fig.~\ref{fig:octant} shows the octant sensitivity for 5+5 years of ($\nu+\bar{\nu}$) running of DUNE. The dark-red solid curve shows the octant sensitivity for the standard case with stable neutrinos. The green dashed curve is for the case when $\nu_3$ is taken as unstable with $\tau_3/m_3=1.2\times 10^{-11}$ s/eV in the data, but in the fit we keep it to be stable. We note that the octant sensitivity of DUNE improves for the green dashed line in the lower octant, but in the higher octant it deteriorates. This is consistent with our observations in Fig.~\ref{fig:th23}. For detailed explanation of this, we refer the reader to the detailed discussion above.

\subsection{CP-violation and Mass Hierarchy Sensitivity}

\begin{figure}
\includegraphics[width=0.5\textwidth]{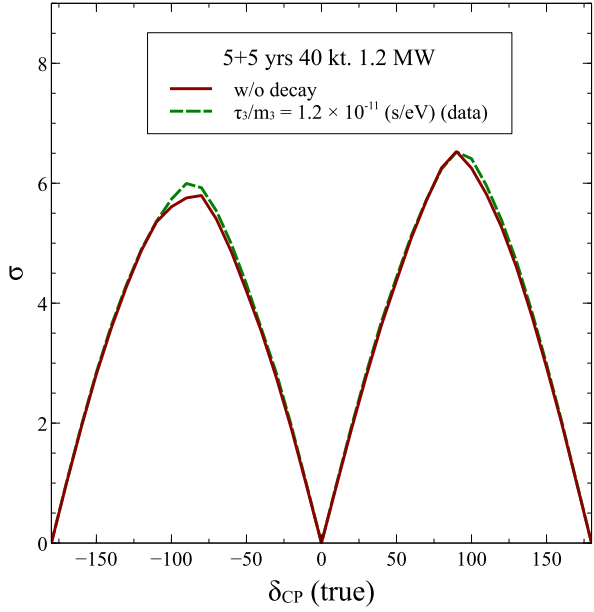}
\caption{\label{fig:cpv}Expected CP-violation sensitivity at DUNE. The dark red solid curve is for standard case of stable neutrinos. The green dashed curve is for the case when $\nu_3$ is taken as unstable with $\tau_3/m_3=1.2\times 10^{-11}$ s/eV in the data, but in the fit we keep it to be stable.}
\end{figure}

\begin{figure}
\includegraphics[width=0.45\textwidth]{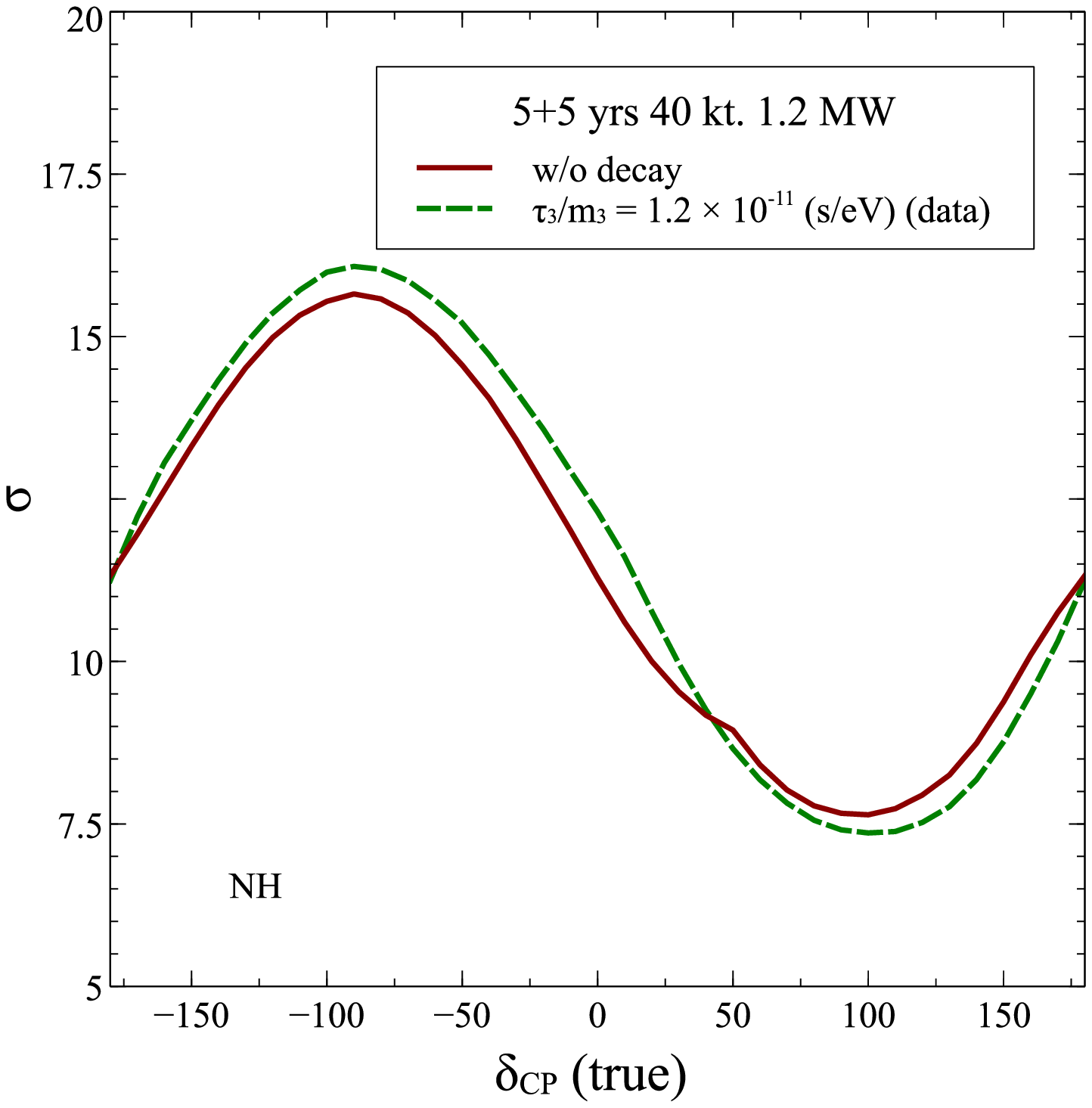}
\includegraphics[width=0.45\textwidth]{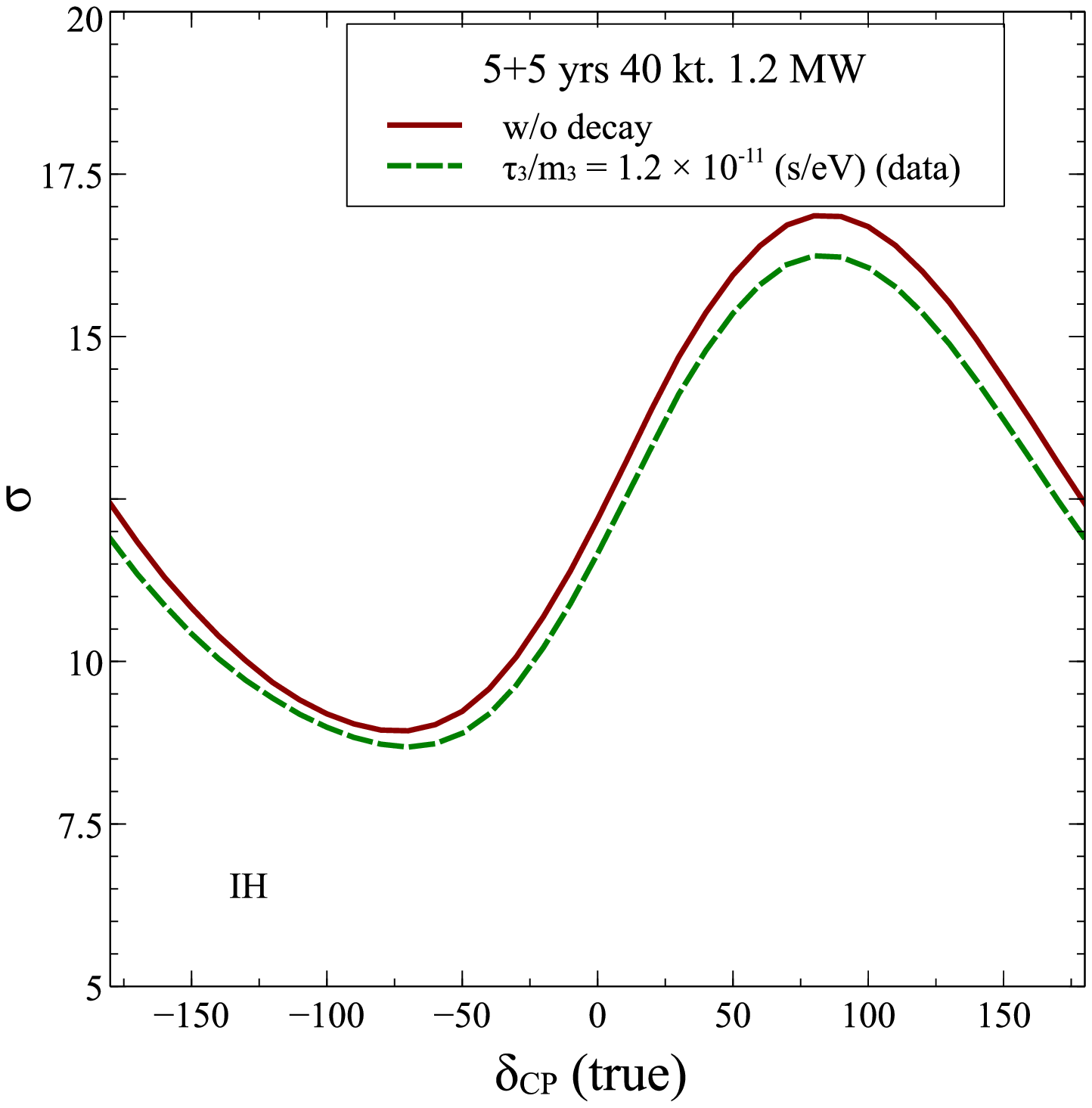}
\caption{\label{fig:mh} Expected mass hierarchy sensitivity at DUNE. The dark red solid curve is for standard case of stable neutrinos. The green dashed curve is for the case when $\nu_3$ is taken as unstable with $\tau_3/m_3=1.2\times 10^{-11}$ s/eV in the data, but in the fit we keep it to be stable. The left panel is for NH true while the right panel  is for IH true. }
\end{figure}

In Fig.~\ref{fig:cpv} we show the expected CP-violation sensitivity at DUNE. As before, the dark red solid curve is for standard case of stable neutrinos. The green dashed curve is for the case when $\nu_3$ is taken as unstable with $\tau_3/m_3=1.2\times 10^{-11}$ s/eV in the data, but in the fit we keep it to be stable. The data was generated at the values of oscillation parameters given in section III and $\theta_{23}=42^\circ$. Decay in the data is seen to bring nearly no change to the CP-violation sensitivity of DUNE, with only a marginal increase in the CP-violation sensitivity seen at $\delta_{CP}$(true)$\simeq \pm 90^\circ$. 

The impact of decay on the expected mass hierarchy sensitivity at DUNE is shown in Fig.~\ref{fig:mh} for both normal hierarchy (NH) true (left panel) and inverted hierarchy (IH) true (right panel ). As in all figures shown so-far,  the dark red solid curve is for standard case of stable neutrinos. The data was generated at the values of oscillation parameters given in section III and $\theta_{23}=42^\circ$. The green dashed curve is for the case when $\nu_3$ is taken as unstable with $\tau_3/m_3=1.2\times 10^{-11}$ s/eV in the data, but in the fit we keep it to be stable. For IH true, the effect of decay in data is to marginally reduce the expected mass hierarchy sensitivity for all values of $\delta_{CP}$(true). The impact for the NH true case is more complicated with the expected sensitivity increasing for some values of $\delta_{CP}$(true) and decreasing for others. However, the net change in the expected sensitivity is seen to be very small compared to the expected mass hierarchy sensitivity at DUNE. Therefore, we conclude that the expected CP-violation sensitivity and mass hierarchy sensitivity at DUNE remain largely unmodified, whether or not neutrinos decay. 

\section{Summary \& Conclusion}

We studied the impact of invisible neutrino decay for the 
DUNE experiment. We assumed that the third mass eigenstate 
is unstable and decays to a very light sterile neutrino. 
The mass of this state $m_4$ is assumed to be smaller than 
the mass of the third mass eigenstate $m_3$ irrespective of 
the hierarchy. We   
did a full three-generation study incorporating 
matter effects in our numerical simulations.  
First, we studied the sensitivity of DUNE 
to constrain the parameter $\tau_3/m_3$ and obtained the expected sensitivity   
$\tau_3/m_3 > 4.50\times 10^{-11}$ s/eV at 90\% C.L. for NH, 5+5 year of 
DUNE data and a 40 kt detector volume. 
This is one order of magnitude improvement over the bound obtained in
\cite{Gomes:2014yua} from combined MINOS and T2K data. 
Of course the bound from T2K and NO$\nu$A is expected  to improve in the future, but here we have concentrated only on the prospective bounds from DUNE.  Note that bound on decay from DUNE is expected to be better than that expected from the full run of current experiments.
We also studied the potential of DUNE to discover  
neutrino decay, should it exists in nature  and found that 
DUNE can discover a decaying neutrino scenario for   
$\tau_3/m_3 > 4.27\times 10^{-11}$ s/eV
at 90\% C.L. with its projected run. 
In addition, we explored how precisely DUNE can constrain the 
decay parameter and showed that for an unstable $\nu_3$ 
with $\tau_3/m_3$ = $1.2 \times 10^{-11}$ s/eV, the no decay 
case gets excluded at $3\sigma$. At 90\% C.L. the allowed range 
corresponding to this true value is given as 
$1.71 \times 10^{-11} > \tau_3/m_3 > 9.29\times 10^{-12}$ in units of s/eV. 

We showed that an interesting correlation exists between the decay lifetime and the 
parameter $\theta_{23}$ both in the appearance probability $P_{\mu e}$ as well as the disappearance probability. 
For values of $\tau_3/m_3$ for which 
fast invisible decay relevant for the baseline under consideration occurs, 
the probability $P_{\mu e}$ decreases. This decrease can be compensated by
a higher value of $\sin^2\theta_{23}$. 
Alternatively, if we assume decay to be present in 
the data then it can be mimicked by a no decay scenario for 
a lower value of $\theta_{23}$ leading to an erroneous determination 
of the latter. Since it is well known that determination of 
$\theta_{23}$ is correlated with the value of $\theta_{13}$, we 
presented contours in the $\theta_{23} - \theta_{13}$ plane 
assuming decay in data and fitting it with a model with no decay. 
We found that the contours show a trend to move towards 
lower $\theta_{23}$ value. The allowed range of $\theta_{13}$ 
also spreads as compared to the only oscillation case, but the 
effect is more drastic for $\theta_{23}$. 

We performed a detailed study of the correlation between decay and $\theta_{23}$ for the disappearance channel $P_{\mu\mu}$ and studied how decay affects the $\theta_{23}$ octant sensitivity in DUNE. Since the position of the minima in both the true and fake octant is determined by the disappearance data while the $\chi^2$ at the fake minima is determined by the appearance data, the effect of decay appears through both channels to affect the octant sensitivity at DUNE and we discussed this in detail. We showed how and why the octant sensitivity of DUNE improved for the lower octant and reduced for the higher octant. 
We also studied the impact of a decaying neutrino on the determination of 
hierarchy and $\dcp$ at DUNE. 
The invisible decay scenario considered in this work affects the 
hierarchy and CP sensitivity of DUNE nominally. 
In conclusion, the DUNE experiment provides an interesting testing ground for 
the invisible neutrino decay hypothesis for  $\tau_3/m_3 \sim 10^{-11}$ s/eV. 

Note Added : While we were finalizing this work, ref. \cite{Coloma:2017zpg} 
came, which also addresses exploration of neutrino decay at DUNE. 
Their emphasis is more on visible decay though they also provide   
a comparison with the invisible decay case. We consider invisible decays to 
light sterile neutrinos and have explored different parameter spaces. 
Hence the two works supplement 
each other. 
We also discussed the impact of decay on the determination of $\theta_{23}$ 
and its octant. In addition we also studied the effect of decay on mass hierarchy  and 
$\dcp$ discovery at DUNE.

\section*{Acknowledgment}
We acknowledge the HRI cluster
computing facility (http://www.hri.res.in/cluster/). SG would like to thank 
Lakshmi. S. Mohan, Chandan Gupta and Subhendra Mohanty for discussions. 
This project has received funding from the European Union's Horizon
2020 research and innovation programme InvisiblesPlus RISE
under the Marie Sklodowska-Curie
grant  agreement  No  690575. This  project  has
received  funding  from  the  European
Union's Horizon  2020  research  and  innovation
programme  Elusives  ITN  under  the 
Marie  Sklodowska-Curie grant agreement No 674896.

\bibliography{ref}

\end{document}

%% file: header.tex
\newcommand{\be}{\begin{eqnarray}}
\newcommand{\ee}{\end{eqnarray}}

\newcommand{\ma}{\Delta m^2_{31}}

\newcommand{\dcp}{\delta_{CP}}

\def\gs{\mathrel{
   \rlap{\raise 0.511ex \hbox{$>$}}{\lower 0.511ex \hbox{$\sim$}}}}
\def\ls{\mathrel{
   \rlap{\raise 0.511ex \hbox{$<$}}{\lower 0.511ex \hbox{$\sim$}}}}
\newcommand{\bea}{\begin{equation} \begin{array}{c}}
\newcommand{\bead}{\begin{equation} \begin{array}{cccc}}
\newcommand{\eea}{ \end{array} \end{equation}}